\documentclass[preprint]{emulateapj}
\slugcomment{Accepted for publication in the Astrophysical Journal}
\shortauthors{Christlein \& Zaritsky}
\shorttitle{Kinematics of Outer Disks of Galaxies}

\begin{document}

\title{The Kinematic Properties of the Extended Disks of Spiral Galaxies: A Sample of Edge-On Galaxies}

\author{Daniel Christlein\altaffilmark{1,2} \& Dennis Zaritsky\altaffilmark{3}}
\shortauthors{Christlein \& Zaritsky}
\altaffiltext{1}{Max-Planck-Institut f\"ur Astrophysik, Garching, Germany}
\altaffiltext{2}{Andes Fellow, U. Chile \& Yale University, Casilla 36-D, Santiago, Chile}
\altaffiltext{3}{Steward Observatory, University of Arizona, 933 North Cherry Avenue, Tucson, AZ 85721}
\email{dchristl@mpa-garching.mpg.de,dzaritsky@as.arizona.edu}

\begin{abstract}         
We present a kinematic study of the outer regions ($R_{25}<R<2\times R_{25}$) of 17 edge-on disk galaxies. Using deep long-slit spectroscopy (flux sensitivity $\sim$ few 10$^{-19}$ erg s$^{-1}$ cm$^{-2}$ arcsec$^{-2}$), we search for H$\alpha$ emission, which must be emitted at these flux levels by any accumulation of hydrogen due to the presence of the extragalactic UV background and any other, local source of UV flux. We present results from the individual galaxy spectra and a stacked composite. We detect H$\alpha$ in many cases well beyond $R_{25}$ and sometimes as far as $2\times R_{25}$. The combination of sensitivity, spatial resultion, and kinematic resolution of this technique thus provides a powerful complement to 21-cm observations. Kinematics in the outer disk are generally disk-like (flat rotation curves, small velocity dispersions) at all radii, and there is no evidence for a change in the velocity dispersion with radius. We place strong limits, few percent, on the existence of counter-rotating gas out to $1.5 R_{25}$. These results suggest that thin disks extend well beyond $R_{25}$; however, we also find a few puzzling anomalies. In ESO 323-G033 we find two emission regions that have velocities close to the systemic velocity rather than the expected rotation velocity. These low relative velocities are unlikely to be simply due to projection effects and so suggest that these regions are not on disk-plane, circular orbits. In MCG-01-31-002 we find emission from gas with a large velocity dispersion that is co-rotating with the inner disk. 
\end{abstract}


\section{Introduction}
\label{sec:intro}

Recent findings promulgate the notion that the baryonic component of spiral galaxies is significantly larger than suggested by classical size indicators such as the $B$-band $R_{25}$ radius (the isophotal radius where the surface brightness drops to 25 mag arcsec$^{-2}$). In an era where the study of dark matter halos has shown spiral galaxies to be at least ten times larger than the optical disk \citep{zsfw,mckay,prada}, one may not be surprised that baryons are also found at large radii, and radio observations of the 21-cm line of neutral hydrogen have anticipated this suspicion by many decades \citep{vdhulst,dieter}. More recently, striking new discoveries are showing that these baryons, which are found in all three principal ``phases" (stars \citep{ferguson2002, ferguson2005,pohlen,zibetti04a, zibetti04,blandhawthorn2005,ellis}, gas \citep{phookun,blandhawthorn,ferguson1998, ferguson1998b,thilker}, and dust \citep{nelson,alton}) are organized in such a way that they are not merely the flotsam left as the forming galaxy falls into the dark matter potential well, but rather the outer extent of the structures manifested at smaller radii. How these structures transition from the baryonic disk, where the baryonic mass has a sharply peaked distribution of specific angular momenta, to the halo of the galaxy, where the measurable baryons, those in globular clusters and satellite galaxies, have a broad distribution of specific angular momenta, is still unclear.

These two realms are not isolated. Hierarchical growth models predict that the galaxy grows by accreting new material from the halo and beyond. As the infalling outer material crashes onto the outer banks of a disk galaxy, it should leave kinematic signatures in the form of lower rotational velocities, increased velocity dispersion, and possibly even counterrotation in the interface region between disk and halo. \citet{vandenbosch} predict that, upon the formation of a galaxy, up to 40\% of the gas in a halo has negative specific angular momentum. If such gas stands any chance of survival, it would be in the outermost regions of late-type galaxies, where dynamical timescales are much larger than in the inner disk, where such signatures in the gas kinematics would be quickly wiped out.

The outer disk may also be a significant repository of positive angular momentum. If the surface mass density of the disk follows an exponential law and the rotation curve remains flat, then around the radius $R_{25}$, the total angular momentum will scale roughly in proportion to the radius. This additional angular momentum in the outer disk sets an even higher bar for hierarchical growth models, which underpredict the angular momentum of galaxies \citep{vandenbosch}. A kinematic inventory of the outskirts of disks is therefore an important aspect of a full accounting of the angular momentum of disk galaxies.

Although radio observations have been the stalwart of this field for
over three decades, and have confirmed that many galaxies are embedded
in extended disks of neutral hydrogen, they have some potentially
important limitations. The large beam size of radio telescopes means
that, at each pointing, material from a wide range of radii,
elongations, and velocities contributes to the observed signal, making
detailed modeling of the emission line necessary. If gas in the
outskirts of galaxies is distributed in clumps rather than smoothly,
such detailed modeling becomes very difficult, and small sources with
anomalous kinematics may be indistinguishable from the velocity
distribution that would be expected across the beam anyway. In
addition, if the gas is ionized due to irradiation by the inner galaxy
\citep{blandhawthorn1998} or the intergalactic ultraviolet background \citep{sunyaev, maloney93,corbelli} then there will be no 21-cm emission. Furthermore, radio interferometric observations require substantial observational efforts to achieve favorable combinations of velocity resolution, spatial resolution, and high sensitivity, and usually strike a trade-off particularly between spatial resolution and sensitivity (for example, \citet{martin1998} lists no arcsecond-scale 21-cm maps of extragalactic objects). Even so, interferometric 21-cm observations for even very nearby galaxies often extend over hours or even days. Finally, 21-cm line observations provide no ancillary information, i.e., other emission lines, that could be used for line diagnostics of density and metallicity.

An underexplored way to study the kinematics of the outer regions of gaseous disks is by detecting optical emission, for example, the H$\alpha$ line. Due to the low level of star formation in the outer disks of galaxies, typical H$\alpha$ rotation curves \citep{vogt,catinella} end well within the optical radius $R_{25}$ of the disk. Nonetheless, H$\alpha$ emission is expected at some level from any accumulation of hydrogen due to the ubiquity of ionizing flux throughout the universe not only from local star formation, but also from nearby galaxies, AGN, and the metagalactic UV background, even though such emission may only occur at extremely faint surface brightness levels ($\sim 10^{-19}$ $erg$ $s^{-1}$ $cm^{-2}$ $arcsec^{-2}$). First attempts to detect such emission were made by \citet{blandhawthorn} with a Fabry-Per\'ot interferometer, revealing outer-disk emission in NGC 253 even beyond the truncation radius of the neutral hydrogen disk at a level of a few $10^{-19}$ $erg$ $s^{-1}$ $cm^{-2}$ $arcsec^{-2}$.  With modern 6-8 m class telescopes and efficient detectors, it is possible to push the detection threshold even with long-slit spectrographs to sufficiently low sensitivities to detect H$\alpha$ well beyond the bright stellar disk, in a realm usually only probed by 21-cm observations, and at the same time make use of the more robust sky subtraction that long-slit spectrography offers over Fabry-Per\'ot observations. Long exposure times (several hours) are required for a spectroscopic detection of the outer disk, but if successful, such detections will provide high spatial and reasonably high kinematic resolution, and a host of ancillary information, such as a measure of the stellar continuum and other emission lines, most notably [SII] and [NII]. 

Most searches for H$\alpha$ emission from the outer disks in the past, including those listed above, have made use of narrow-band imaging. However, for line emission, optical spectroscopy is capable of achieving a much higher S/N per pixel than imaging, as it allows us to isolate the line flux from the sky and stellar continuum background much more reliably. We can therefore probe much fainter structures in the outer disk, including not only selected bright HII regions, but also the diffuse interstellar medium.

We report here on the first results of an observational campaign to study the outer disks of galaxies with long-slit spectroscopy. Specifically, in this paper, we discuss constraints on the kinematics of the outer disk. In \S \ref{sec_theory}, we discuss what we might detect in H$\alpha$ in the outer disks. In \S \ref{sec_data}, we describe our observational campaign. In \S \ref{sec_analysis}, we outline our reduction and analysis pipeline. Our results and discussions thereof are presented in \S \ref{sec_results}: We examine rotation curves out to radii of $2\times R_{25}$, i.e., our best cases are competitive with the extent reached by radio interferometric studies. We specifically search for kinematic anomalies, i.e., gas that is systematically deviating from co-rotation, taking advantage of the high spatial resolution of optical spectroscopy. We also discuss what constraints can be placed on the profile of a potential dark matter halo at large radii from these data. In addition, we use a stacked composite spectrum to examine the velocity dispersion of gas in the outer disk and to search for gas on counterrotating orbits. 

Subsequent papers will discuss the physical state of the outer-disk gas, including metallicity and density and constraints on the intergalactic UV background, and will examine the relation between H$\alpha$ emission and stellar populations in the outer disk.

\section{Observations}

\label{sec_data}

\subsection{Sources of H$\alpha$ emission in the outer disk}
\label{sec_theory}

The brightest and most easily detected source of H$\alpha$ emission are local star formation regions in the outer disk. Such regions have been identified using narrow-band imaging \citep{ferguson1998} and examined spectroscopically \citep{ferguson1998b}.  The luminosities of these objects are in the range of $0.1-8 \times 10^{38}$ erg s$^{-1}$, although the lower limit is observational rather than physical. Using a long-slit spectrograph on an 8-m class telescope for a galaxy at a cosmological redshift of cz=4,000 km s$^{-1}$, such star formation regions are typically detectable within several minutes. However, these objects highlight only small, selected regions, and may not cover the full extent of the gas distribution. Furthermore, in a blind survey with a long-slit spectrograph, it is a matter of chance whether they fall onto the slit.

A more comprehensive study of the baryon budget is possible if the
outer disk is irradiated uniformly by a sufficiently strong exterior
ionizing flux. The inner, star-forming disk itself can constitute such
a source, especially if the disk is slightly warped \citep{blandhawthorn1998}. The H$\alpha$ surface brightness in this case depends strongly on the star formation rate in the inner disk, the UV escape fraction, and the geometry of the disk, so the necessary exposure times can vary between minutes and several hours. A second such source is the intergalactic UV background. The intensity of this radiation field is still uncertain; a theoretical prediction estimates $1.3^{+0.8}_{-0.5}\times 10^{-23}$ erg s$^{-1}$ cm$^{-2}$ sr$^{-1}$ Hz$^{-1}$\citep{shull99}. \citet{vogel95} provide an upper limit of $<8\times 10^{-23}$ erg s$^{-1}$ cm$^{-2}$ sr$^{-1}$ Hz$^{-1}$, and a recent measurement from \citet{scott02} finds $7.6^{+9.4}_{-3.0}\times10^{-23}$ erg s$^{-1}$ cm$^{-2}$ sr$^{-1}$ Hz$^{-1}$. Ionization by the intergalactic field sets the lowest surface brightness limit of H$\alpha$ emission, but even it is within the reach of large telescopes and modern instruments. Because H$\alpha$ emission stimulated by the UV background provides the most uniform and unbiased probe of baryonic material --- wherever there is hydrogen, there must be some H$\alpha$ emission ---, we set our exposure times to provide a realistic chance of detecting background-ionized hydrogen for at least some of our target galaxies. The H$\alpha$ surface brightness level corresponding to the background flux of \citet{scott02} is $\sim 2^{+3}_{-1} \times 10^{-19}$ erg s$^{-1}$ cm$^{-2}$ arcsec$^{-2}$ (for derivations, see, e.g., \citet{maloney93, blandhawthorn}). This sensitivity is attainable with an 8-m class telescope within several hours of exposure time.

\subsection{Target Selection}

We select target galaxies using the following criteria: 1) visibility --- targets should be observable during the entire night and have the lowest possible mean airmass, 2) redshift --- targets should have recessional velocities between 3,600 and 12,000 km s$^{-1}$ for which the redshifted H$\alpha$ line falls into a dark and flat region of the night sky spectrum, 3) angular size --- the slit length should cover the region of interest, ideally the entire galaxy, and leave room for blank sky, 4) morphology --- targets should be predominantly late-type galaxies (Sc-Sd) because these are generally gas-rich, 5) H I content --- targets are preferred if they have existing detections in the literature, 6) inclination --- targets must be close to edge-on configuration to maximize the chance that disk regions will lie within the slit, and 7) environment --- targets must be relatively isolated both in regards to possible galaxy companions, but also in terms of bright stars that would affect the observations.

These criteria are somewhat flexible, and in particular we do include an Sb, an Sab, and a gas-rich S0 (which will be discussed in a separate paper; Kannappan et al., in prep.). Our sample does not constitute a complete or systematic sample in any way, but we emphasize that, with the exception of the S0, all of them were selected for appearing as undisturbed as possible. In practice, the axis ratios $(b/a)$ of our target galaxies are $\leq 0.2$, with the exception of one object --- UGC 10972, for which it is 0.27. The inclination angle in Table \ref{tab_galdata} is calculated directly from the published axis ratio, neglecting the finite thickness of the disk, and is typically around 80 degrees. If more realistic values for the thickness are adopted (i.e., $c/a\leq 0.2$) all galaxies with the exception of UGC 10972 are consistent with exact edge-on configuration.  This becomes an important distinction, because in a perfectly edge-on configuration, the line of sight traverses the entire disk and thus a wide range of line-of-sight velocities, leading to possible ambiguities in the interpretation of the kinematic measurement. In a slightly inclined system, the slit probes a much smaller range of depths. We will discuss this topic more extensively below in connection with measurement uncertainties and biases. 

The selected galaxies are listed in Table \ref{tab_galdata}, along with their redshift, the isophotal radius $R_{25}$, the inclination angle, morphological type, and approximate absolute $B$-band magnitude, as gleaned from the NASA-IPAC Extragalactic Database (NED) and calculated assuming a $\Lambda$CDM cosmology with H$_{0}$=71 km s$^{-1}$ Mpc$^{-1}$ and $\Omega_{m}$=0.27, $\Omega_{\Lambda}$=0.73, and $\Omega_{k}$=0.

\begin{deluxetable*}{lrrrrrlr}
\tablecaption{Basic Galaxy Data}
\tablewidth{0pt}
\tablehead{
\colhead{Galaxy}&
\colhead{cz}&
\colhead{R$_{25}$}&
\colhead{R$_{25}$}&
\colhead{b/a} &
\colhead{i} &
\colhead{Type} &
\colhead{$M_{B}$}\\
&
[km s$^{-1}$]&
[arcsec]&
[kpc]&
&
[$^\circ$]
&
&
}
\startdata
ESO 201-G022         & 4068 &  75.7 &  20.7 & 0.14 & 82 & Sbc HII    & $-$19.0 \\
ESO 299-G018         & 4821 &  61.5 &  19.9 & 0.11 & 84 & Sc         & $-$18.4 \\
ESO 478-G011         & 5216 &  42.1 &  14.7 & 0.12 & 83 & Scd        & $-$18.1 \\
IC 2058              & 1379 &  93.1 &   8.6 & 0.15 & 81 & Sc         & $-$17.5 \\

ESO 323-G033         & 2348 &  58.7 &   9.3 & 0.16 & 81 & Sd         & $-$17.7 \\
ESO 380-G023         & 2753 &  41.6 &   7.7 & 0.18 & 80 & Scd        & $-$17.3 \\
ESO 385-G008         & 3756 &  62.2 &  15.7 & 0.10 & 84 & Sd         & $-$18.3 \\
IC 4393              & 2753 &  80.2 &  14.9 & 0.11 & 84 & Scd        & $-$18.5 \\

MCG -01-31-002       & 5741 &  42.4 &  16.2 & 0.17 & 80 & Scd        & $-$19.6\\
ESO 445-G081         & 4352 &  60.8 &  17.8 & 0.18 & 80 & Sbc        & $-$19.8 \\
ESO 445-G085         & 4285 &  57.4 &  17.0 & 0.06 & 86 & Scd        & $-$17.7 \\
ESO 568-G010         & 5507 &  52.4 &  19.3 & 0.06 & 87 & Sd         & $-$18.1 \\
ESO 381-G045         & 7129 &  33.0 &  15.6 & 0.20 & 78 & Sb         & $-$19.6 \\

UGC 09138            & 4601 &  58.5 &  18.1 & 0.12 & 83 & Sc         & $-$18.9 \\
UGC 09780            & 5173 &  62.7 &  21.7 & 0.12 & 83 & Scd        & $-$19.0 \\
UGC 10453            & 4352 &  41.4 &  12.1 & 0.10 & 84 & Sd         & $-$18.2 \\
UGC 10972            & 4652 &  75.3 &  23.5 & 0.27 & 74 & Scd        & $-$20.0 \\
\enddata
\label{tab_galdata}
\tablecomments{Data are drawn from NED. Isophotal radii R$_{25}$ in the $B$-band are taken from the ESO-LV catalog \citep{esolv} for all ESO and IC galaxies, and from the RC3 D25 catalog \citep{devauc} for MCG -01-31-002 and the UGC galaxies. Inclination angles are calculated assuming the idealized case of an infinitesimally thin disk.}
\end{deluxetable*}

\subsection{Instruments}

Our observational campaign uses long slit spectrographs. Their advantages for our program include an extended field of view that covers the entire major axis of a galaxy simultaneously plus a substantial sky background, and sufficient velocity resolution to determine rotation curves.

A moderate dispersion spectrograph (of the order 1\AA/pixel) is suitable for this project. High dispersion spectroscopy will improve the kinematic precision, but dilute the H$\alpha$ line and so increase the relative contribution of read noise. Low resolution spectroscopy would decrease the kinematic precision below what is necessary to measure kinematics internal to the galaxy; furthermore, since the width of the line image on the detector is set primarily by the slit width, low-resolution spectroscopy would increase the contribution of sky across this width. Anamorphic demagnification, which exists in some of the spectrographs used for our program (the B\&C and GMOS-S spectrographs) further aids our observations by reducing the effective slit image width, and thus minimizes noise contributions from read noise and sky background.

We observed using the following instruments: a) the Boller \& Chivens spectrograph on the Magellan I (Baade) telescope using the 1200 l/mm grating blazed at 5500 \AA\ with a 1 arcsec slit on the nights of 26-28 Mar 2003 and 16-17 Apr 2004, b) the Red Channel Spectrograph on the MMT using the 1200 l/mm grating blazed at 7700 \AA\ with a 1 arcsec slit on the nights of 21-22 May 2004, c) FORS-1 on the VLT-2 with a 1 arcsec slit and the 600R grism with a dispersion of 1.038 \AA pixel$^{-1}$ (unbinned) on Nov 16-18, 2004, and d) GMOS on Gemini-South with a 1 arcsec slit and the R831 grating with a dispersion of 0.34 {\AA} per unbinned pixel on Apr 6-8, 2005. Aside from the througput and sensitivity differences, the principal difference in the setup between the GMOS and FORS1 observations on the one hand and the B\&C and Red Channel observations on the other hand lies in the smaller slit length available on the latter two spectrographs. In these cases, we first obtained a short ($\sim 5-10$ min) spectrum with the galaxy centered on the slit, followed by much longer exposures with an offset along the major axis, usually in only one direction.

Individual exposures, typically 20-30 min in length on the MMT and Magellan, and one hour on the VLT and Gemini, are combined to result in the total observing time given in Table \ref{tab_obs}. Additional details of the observing runs are also provided in Table \ref{tab_obs}. 

\begin{deluxetable}{lrrrr}
\tablecaption{Summary of Observing Runs}
\tablewidth{0pt}
\tablehead{
\colhead{Galaxy}&
\colhead{Offset}&
\colhead{Telescope}&
\colhead{Date}&
\colhead{t$_{EXP}$}  \\
&&&&[sec]
}
\startdata
ESO 201-G022 & central & VLT/FORS1 & Nov 2004 & 37,329 \\
IC 2058      & central & VLT/FORS1 & Nov 2004 & 12,399 \\
ESO 299-G018 & central & VLT/FORS1 & Nov 2004 & 12,199 \\
ESO 478-G011 & central & VLT/FORS1 & Nov 2004 & 10,300 \\

ESO 323-G033 & central & Gemini-S/GMOS & Apr 2005 & 39,600 \\
IC 4393      & central & Gemini-S/GMOS & Apr 2005 & 15,400 \\
ESO 380-G023 & central & Gemini-S/GMOS & Apr 2005 & 10,800 \\
ESO 385-G008 & central & Gemini-S/GMOS & Apr 2005 &  7,200 \\

ESO 381-G045 & NE      & MAG/B\&C& Mar 2003 & 50,400\\
ESO 385-G008 & NE      & MAG/B\&C           &  Apr 2004    &4,800\\
ESO 445-G081 & NE      & MAG/B\&C           &  Apr 2004    &19,200\\
ESO 445-G085&NE&MAG/B\&C& Apr 2004&18,600\\
ESO 568-G010&NE&MAG/B\&C& Apr 2004&15,800\\
IC 4469&NW&MMT/Red& May 2004&8,284\\
MCG-01-31-002&NE&MAG/B\&C& Apr 2004&9,000\\
UGC 09138&NW&MMT/Red& May 2005&3,600\\
UGC 09780&SE&MMT/Red& May 2004&3,600\\
UGC 09780&NW&MMT/Red&May 2004&4,800\\
UGC 10453&SW&MMT/Red& May 2005&4,800\\
UGC 10972&NE&MMT/Red&May 2005&4,800\\
UGC 10972&SW&MMT/Red&May 2005&3,600\\
\enddata
\label{tab_obs}
\tablecomments{``Offset'' indicates where the slit center was positioned relative to the galaxy (offsets are always along the major axis).}
\end{deluxetable}

\section{Data Analysis}
\label{sec_analysis}
\subsection{Reduction}
We reduce the long-slit spectra from the MMT, Magellan, and the VLT with the standard IRAF reduction tasks. For Gemini data, we use the GEMINI package in version 1.7\footnote{During the reduction, we discovered an error in the GSFLAT algorithm that caused the task to carry out the flatfielding with an individual flatfield, instead of a combined master flatfield; this problem was quickly patched by the Gemini observatory staff, and we continued the reduction with the patched version of the package.} For the Gemini and VLT data, we determine wavelength solutions from arc spectra obtained immediately before or after the science exposures. If there are any indications that the wavelength solution changed between exposures, we use separate solutions for different images or groups of images. For the Magellan and MMT spectra, we use sky lines to wavelength calibrate. We sky subtract using the sky spectrum from the ends of the slit. The region that we consider to be devoid of emission from the galaxy itself is determined by eye, and typically corresponds to a fraction of at least half the slit length. We continuum subtract by fitting low-order polynomial functions along the dispersion axis.

\subsection{Measurements}

\label{sec_measure}

We use the H$\alpha$ line for the measurements of the gas kinematics in the outer disk. Although other lines ([NII], [SII]) are observed, these are usually very weak in the outer regions of our target galaxies. To detect and extract the H$\alpha$ line, we begin by applying a smoothing filter with a width of $\sim 1.5$ arcsec in the spatial direction to the spectrum. This scale is about twice the typical seeing and is chosen as a compromise to detect both small-scale features and attain sufficient sensitivity. Starting at a local maximum near the wavelength range where H$\alpha$ emission is expected, we define a search window whose width along the dispersion axis narrowly encloses the entire line, and determine the mean intensity-weighted wavelength within this window as the new estimate for the line centroid. We iterate this process, setting the new estimate for the centroid as the new center of the search window. This process is carried out interactively, using software that provides a direct visual feedback of the line extraction process and allows us to correct errors. Failures in the automatic centroid search are corrected manually by restarting the iteration near the correct position.  At very large radii, where visual inspection often yields no significant peak, the decision where to start the iteration admittedly becomes arbitrary, but in such cases the resulting detections generally turn out to be below the significance threshold as well; i.e., the risk of selecting spurious peaks due to ``hunting for significance'', is not a major problem in practice. After convergence, fluxes are determined by integrating the emission within the search window.

A source of systematic error in the determination of fluxes, and thus of the significance of a detection, is broad H$\alpha$ absorption by the underlying stellar population. In the bright inner disks, H$\alpha$ absorption is negligible, compared to the H$\alpha$ emission, but in some cases towards the outer edge of the disk, where the stellar continuum is noticeably more extended than H$\alpha$ emission, it can lead to a systematic underestimate of the H$\alpha$ flux, or even suppress the H$\alpha$ emission completely. To remove this local background, we define regions along the dispersion axis between the H$\alpha$ emission line and the [NII]6548 and [NII]6584 lines, where the H$\alpha$ absorption is observed, at fixed wavelength offsets relative to the H$\alpha$ centroid. We then fit --- for each row of the input spectrum --- a Gaussian to the H$\alpha$ absorption trough in these local background regions. Typically, the Gaussian parameters (width, amplitude, centroid) exhibit large scatter from one row to the next. In a next step, we therefore fit the parameters of the individual Gaussian fits as a function of the position along the slit with simple low-order functions (an {\it arctan} function for the centroid, Gaussian for the width, and Gaussian or exponential, as suggested by the data, for the amplitude). With the parameters interpolated from these fits, we model the local H$\alpha$ absorption trough in each row. We also review the quality of the local background model and apply manual corrections on a row-by-row basis where necessary. In particular, we adopt a conservative approach towards outlying emission regions, and in general do not accept an isolated detection at large distances from the galaxy as real if it is pushed over the significance threshold only by the local background subtraction (unless the presence of H$\alpha$ absorption is unmistakable).

\subsection{Uncertainties}
\label{sec_uncertainties}
Uncertainties in the measured flux are determined in two ways: First, we apply a bootstrapping algorithm, which randomply resamples the CCD rows within the area covered by our measurement window, and measures the flux from this random sample. By repeating this and determining the standard deviation of the results, we obtain a first estimate for the flux uncertainty. While this algorithm accounts well for variations on pixel-to-pixel scales, including Poisson errors in the H$\alpha$ flux, it is insensitive to fluctuations in the sky or continuum background that occur on scales comparable to that of the measurement window or larger. Such fluctuations frequently result from flatfielding imperfections and ultimately a more severe limitation of our ability to detect faint H$\alpha$ emission than the Poisson errors in the sky background. In order to quantify the uncertainties in the sky background, we define several control regions along the dispersion axis that do not contain prominent emission or absorption lines. We then slide our measurement window across this background region, determining the integrated flux at each position. Since these control regions are typically at least an order of magnitude larger than the measurement window, the standard deviation of the measured residual fluxes around their mean is a good measure of the typical errors in the sky background within the measurement window. As our final flux uncertainty, we adopt the larger of the two estimates described in this paragraph. Usually, the uncertainties determined after the first method are larger in the central parts of the galaxy, where the Poisson errors in the H$\alpha$ emission dominate, while at large radii, the large-scale fluctuations in the sky background are the dominating source of error.

\begin{figure}[t]
%\plotone{eso201.errors.eps}
\plotone{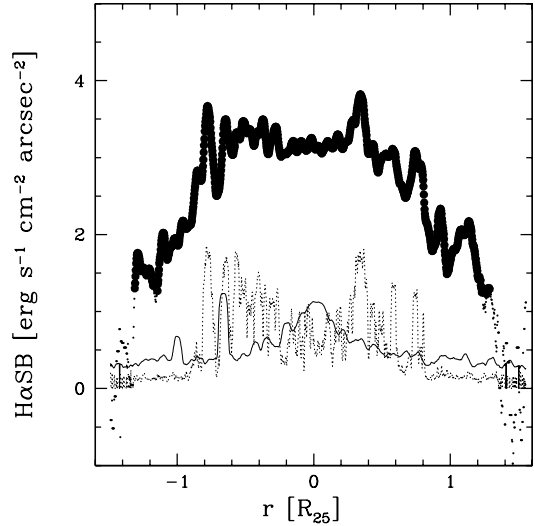}
\caption{Comparison of the two uncertainty estimates in the case of ESO 201-G022. The horizontal axis shows the distance from the galaxy center in units of $R_{25}$. The vertical axis shows the H$\alpha$ surface brightness and is linear below a value of 1, and decadically logarithmic above (i.e., an abscissa of 2 signifies a value of 10 etc.). Bold data points show significant flux detections, small dots insignificant ones. The dotted line indicates flux uncertainties from bootstrapping, and the solid line shows background uncertainties. In the inner disk, uncertainties based on the bootstrapping method, which include Poisson errors from H$\alpha$ emission and the continuum, dominate, but the outer disk is dominated by errors in the background, which includes not only Poisson noise in the sky background, but also correlated errors, such as flatfielding imperfections. We use the larger of the two to evaluate the significance of a detection.}
\label{fig_errors}
\end{figure}

In Fig. \ref{fig_errors}, we show how the flux uncertainty varies as a function of radius, using one of our deepest and most extended spectra (ESO 201-G022) as an example. Bold data points mark flux measurements that we consider significant, and small dots denote insignificant measurements. The dotted, thin line at the bottom represents the flux uncertainties as determined by the first method (bootstrapping within the measurement window), while the bolder, continuous line shows the background flux uncertainties, obtained by the second method described above. The plot is linear below an ordinate of 1, and decadically logarithmic above (i.e., a value of 1 represents a measurement of 1 of the measured units, while a value of 2 represents 10 units, and a value of 0.5 represents 0.5 units).

 For each spectrum, we determine a detection threshold by fitting a low-order functional form to the extracted data points and then plotting the residual velocity versus the fit as a function of the significance of the measured flux. In order to maintain a conservative approach, this detection threshold may vary from spectrum to spectrum, depending on the amount of flatfielding imperfections, skyline residuals, and other factors, and is between 3 and 7 $\sigma$ for our sample.

Uncertainties in the velocity centroid are determined analogously to the first method, i.e., we measure the standard deviation in the recovered line centroids as we apply a bootstrapping algorithm that randomly resamples the CCD rows within our measurement window (the intensity distribution along the dispersion axis is not affected by the bootstrapping) and estimates the dispersion of the recovered centroids.

This method reflects only the statistical uncertainty in the centroid position, not in the actual line-of-sight velocity. In practice, another source of error dominates: if the slit is not uniformly or at least symmetrically illuminated, the recovered line centroid will not reflect the actual line-of-sight velocity, and surface brightness fluctuations across the slit width may be misinterpreted as velocity shifts. With the spectrographs used in our study (dispersion and slit width), this effect can theoretically introduce deviations in the recovered velocity of several tens of km s$^{-1}$. Higher dispersion, narrow slits, and stronger anamorphic demagnification of the slit image, but also larger seeing, help reduce this effect. In general, however, this issue is only a problem in determining the velocities of individual emission regions. Due to the small angular width of the slit relative to the angular sizes of the galaxies, sources are distributed stochastically across the slit, so that, for large samples --- i.e., those binning over a large spatial extent, or those co-adding multiple galaxies ---, this effect will not introduce strong systematic deviations in the rotation curve. We quantify the effect on the line centroid measurements of individual regions by examining the scatter of the measured line centroids about the fitted rotation curve. The measured scatter is on the order of 10-20 km s$^{-1}$. This result does not preclude the possibility that actual velocity fluctuations with such amplitudes may exist, but we do not attempt to distinguish between these two possibilities in this analysis; such a distinction would only be possible via the shape of the line profile, which is only viable in the high-surface brightness inner regions of the galaxy.

In addition to this effect, which introduces a random uncertainty in the recovered line-of-sight velocity, there is a potential systematic source of error. Deriving the line-of-sight velocity from the intensity-weighted mean wavelength, as we have done, is an approximation. In reality, any line of sight traversing an extended part of a strongly inclinded disk will intercept parcels of emitting gas at a range of line-of-sight velocities, which are not distributed symmetrically with respect to the rotational velocity at the projected radial distance. This usually leads to an underestimation of the true rotational velocity by generating a tail in the emission profile towards lower velocities. For geometric reasons, the effect is stronger near the center of the galaxy than on the outskirts (it depends primarily on the ratio of the slit width to the projected distance from the center of rotation). This effect is a serious concern in radio astronomy, where beam widths are large and velocity differentials across the beam can be significant. With optical spectroscopy, it is generally smaller by an order of magnitude, because the width of the spectrograph slit is small compared to the typical diameter of a galaxy. To quantify this bias, we have estimated the ratio of the recovered mean rotational velocity to the true rotational velocity, using simple models with a constant rotational velocity and gas surface density, as well as finite thickness of the gaseous disk. For an inclination angle of 85$^{\circ}$, the deviation between the mean flux-weighted velocity, which we recover, and the true rotational velocity is within our typical uncertainties as long as the ratio of the projected radial distance to the slit width is greater than 30. This is the case at the R$_{25}$ radius for all galaxies in our sample. For all other galaxies, the effect becomes a concern at R$_{25}$ only for inclination angles of 88$^{\circ}$ or more. This is a conservative estimate; in reality, the H$\alpha$ flux drops steeply at large radii, so that the flux in the low-velocity tail is further suppressed. Final affirmation that this effect is no great concern (even though most of our sample is consistent with exact edge-on confirmation) comes from the data themselves: In our composite spectrum (which we discuss in detail in \S \ref{sec_composite}), the H$\alpha$ line appears symmetric which no indication of a strong low-velocity tail at any radius.

In edge-on galaxies, dust is known to affect optical rotation curves: At small projected distances from the nucleus, dust extinction suppresses emission from the inner parts of the galaxy, increasing the relative contribution of emission from larger radii along the same line of sight, which tends to have a smaller line-of-sight velocity; in other words, the low-velocity tail discussed in the previous paragraph is emphasized relative to the peak of emission near the rotational velocity. This is known to severely bias the slope of the rotation curve at small radii \citep{giovanelli}. For this reason, our survey is not ideally designed for determining the shape of the rotation curve at small radii. At larger radii, however, all these biasing effects are weaker: not only is the velocity gradient within the field of view smaller, as argued above, but we also expect dust absorption to be a very minor problem around R$_{25}$ and beyond, where the focus of our study lies (which is also noted by Giovanelli \& Haynes).

An additional source of systematic errors in optical spectra arises from possible misalignment between the slit and the galaxy nucleus; however, this effect is only a concern at extremely small radii. Since the present investigation does not focus on the shape of the rotation curve near the nucleus, we neglect this effect.

Finally, if there are strong warps present in the outer disk and their
line of nodes is aligned perpendicular to the line of sight, then the
true inclination angle of the outer gaseous disk may be different from
that derived from the optical axis ratio, and the observed
line-of-sight velocity may be an underestimate of the true rotational
velocity in the warped disk. This possibility cannot be ruled out from
optical data along a one-dimensional slit alone. However, in our data,
we do not see the signature of such an effect (a sudden drop in the
recovered line-of-sight velocity), which may indicate that either such
configurations (i.e., warp tilts on the order of 25$^{\circ}$ or more,
required to introduce a significant dropoff in the velocity) are rare,
or that our rotation curves do not extend to such radii. The possibility exists that an actual rise in the rotation curve is cancelled out exactly by this tilting effect, but it is highly unlikely that these would conspire to reproduce an undisturbed-looking rotation curve in every single case of our sample.

\subsubsection{Flux Calibration}

We flux-calibrate our spectra using the standard stars GD108 and LTT9491 for our VLT data from November 2004, and LTT6248 for the Gemini-S data from April 2005. We calculate fluxes by integrating over the entire line width, and multiplying with the calibration factor (which is measured in erg cm$^{-2}$ count$^{-1}$). We then convert the flux into a surface brightness, using the slit width (1 arcsec for all galaxies in this sample), equivalent exposure time, and platescale in the spatial direction. Flux calibrations from any one run are consistent to within 5-10\%.

For data from Magellan and the MMT, no spectrophotometric standard star observations are available. We use the one object (ESO 385-G008) observed both with Gemini-South and Magellan, to derive an approximate relative calibration factor between the two optical systems, and determine that factor to be $\sim 4.11$ (i.e., for the same incident flux, Gemini-S/GMOS measure a count rate 4.11 times higher). We assume the same value for the MMT (the diameters of the MMT and Magellan primary mirrors are comparable; however, the Red Channel spectrograph on the MMT is likely to be more sensitive than the B\&C spectrograph on Magellan, therefore this assumption is very approximate). We are aware of the inherent uncertainties in this relative calibration, but in this paper, we only use the absolute flux to assign a statistical weight to each spectrum when superimposing data from different runs and/or objects. Therefore, the actual impact of the calibration on our results is minimal.

\section{Results}
\label{sec_results}
\subsection{Rotation Curves}
\label{sec_rcs}

The rotation curve at large radii is a kinematic probe of the mass profile in a region where luminous matter contributes little to the total mass budget. Steep declines in the rotation curve would therefore indicate a break or truncation in the dark matter halo density profile. Although the general existence of massive dark matter halos at radii much larger than those probed here is inferred by studies of satellite galaxies \citep{zsfw} and weak lensing \citep{brainerd}, these results are statistical in nature. H$\alpha$ rotation curves, along with 21-cm rotation curves, are the sole avenue to study the halo mass profiles of a large sample of individual spiral galaxies, albeit at radii smaller than those probed by satellite galaxies and lensing. Furthermore, kinematic anomalies, such as an increased velocity dispersion, or individual emission regions deviating from co-rotation, are possible signatures of the hierarchical build-up of a galaxy predicted under the current cosmological paradigm. At large radii, where dynamical timescales are long, such signatures are preserved for longer periods than in the inner disk.

In this section, we present our results for the rotation curves of the individual galaxies in our sample, based on the measured H$\alpha$ velocity centroids. Our sample is designed to constrain the rotation curve at relatively large radii, close to and beyond the isophotal radius $R_{25}$. Stellar continuum emission \citep{pohlen} and, in fact, resolved stellar populations in local group galaxies \citep{ferguson2002} have been measured to signifianctly fainter surface brightnesses, and $R_{25}$ therefore certainly does not represent a strict cut-off radius of the stellar disk. Nonetheless, this radius is easily reproducible and contains $\sim 90 \%$ of the total flux of a galaxy. Typically, $R_{25}$ is approximately twice as large as the half-light radius, and 5-10 times the surface brightness scale length in the $B$-band. It is therefore a very suitable measure of the radius of the bright, stellar disk of a galaxy, and we cast our measurements of radial distances in units of $R_{25}$ to facilitate the comparison between different galaxies.

The values of $R_{25}$ used here are drawn from the NASA Extragalactic Database (NED) and come from two sources: the Surface Photometry Catalogue of the ESO-Uppsala Galaxies \citep{esolv} and the Third Reference Catalog of Bright Galaxies \citep{devauc}. Table \ref{tab_galdata} indicates which values have been drawn from which source. Where available, we prefer the values of \citet{esolv}, except in the case of ESO 323-G033, where, judging from inspection both of the Digitized Sky Survey plates and our acquisition images, we consider the value of \citet{devauc} to be more appropriate.

While $R_{25}$ is a reproducible measure of the radius of a galaxy disk, some caution must be exercised when comparing measurements in units of $R_{25}$ between the present work and other studies of the outskirts of galaxies \citep{ferguson1998,pohlen}, which generally focus on face-on galaxies. In the absence of dust (an idealistic assumption, but more justified for the outskirts of galaxies than for the gas-rich inner disk), simple geometrical considerations lead us to expect the surface brightness along the line of sight to vary roughly as $\sim 1/cos(i)$, where $i$ is the inclination angle. Therefore, a given isophotal radius will lie $\sim 2.5$ exponential scale lengths further out when the galaxy is seen edge-on than when it is seen face on. Because $R_{25}$ typically corresponds to anywhere between 5 and 10 exponential scale lengths, $R_{25}$ could change by 25 to 50\% due to inclination. Therefore, quantitative comparisons of the distribution of outer-disk material in edge-on and face-on galaxies are complicated.

\begin{figure}
%\plotone{fitrc3.eps}
\plotone{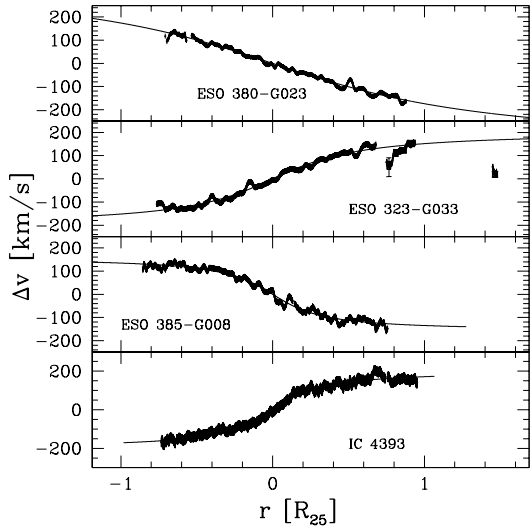}
\caption{Velocity centroids of H$\alpha$ measured relative to the systemic galaxy velocity on a row-by-row basis in the two-dimensional spectrum. The galaxies in this figure were observed with Gemini-S/GMOS. The solid line represents the best {\it arctan} fit. Open symbols show measurements where, to improve S/N, we have binned over larger regions of interest.
}
\label{fig_bdcomp}
\end{figure}

\begin{figure}
%\plotone{fitrc4.eps}
\plotone{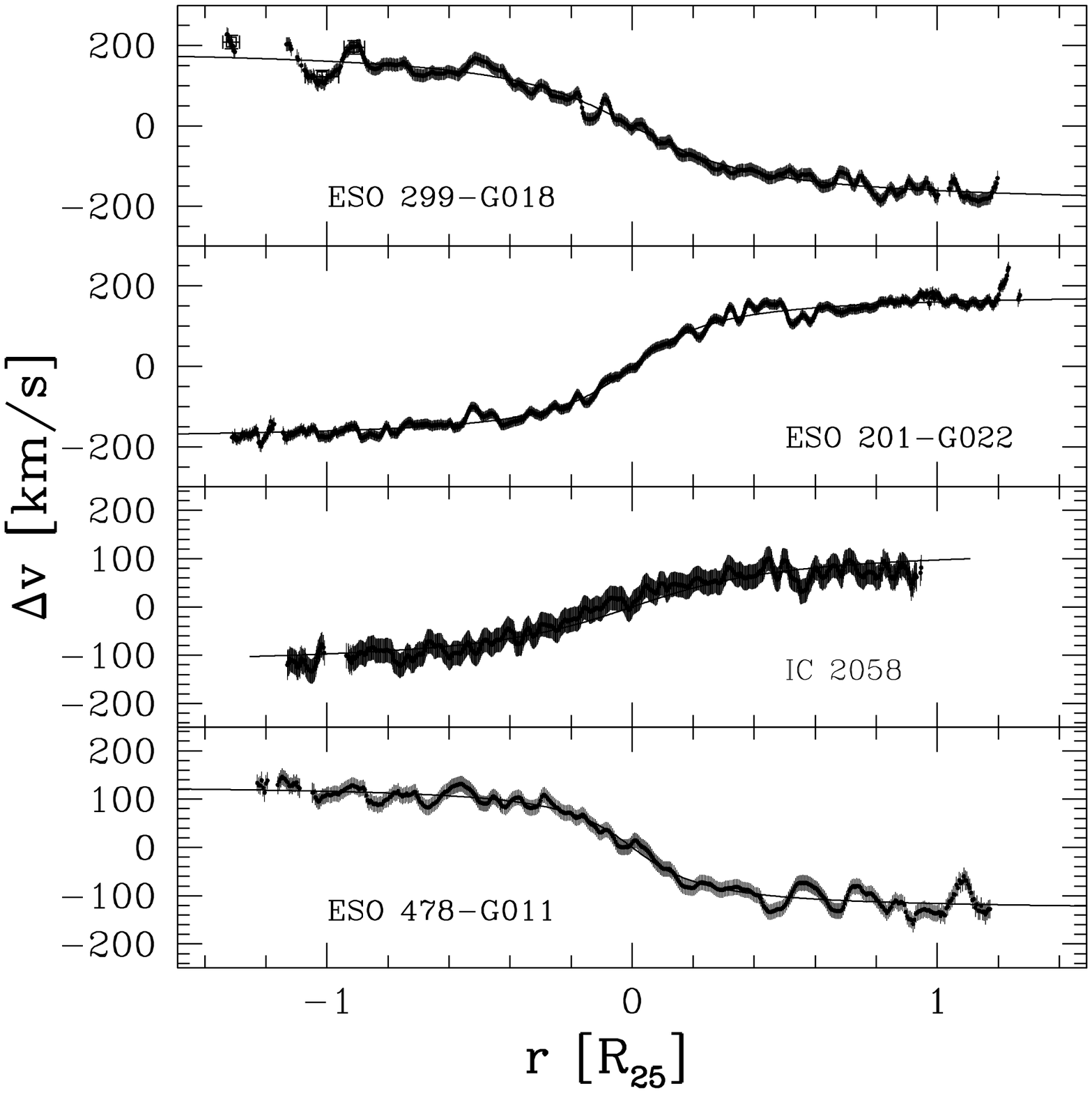}
\caption{Velocity centroids of H$\alpha$ measured relative to the systemic galaxy velocity on a row-by-row basis in the two-dimensional spectrum. The galaxies in this figure were observed with VLT/FORS1. The solid line represents the best {\it arctan} fit. Open symbols show measurements where, to improve S/N, we have binned over larger regions of interest.
}
\label{fig_bdcomp2}
\end{figure}

\begin{figure}
%\plotone{fitrc.eps}
\plotone{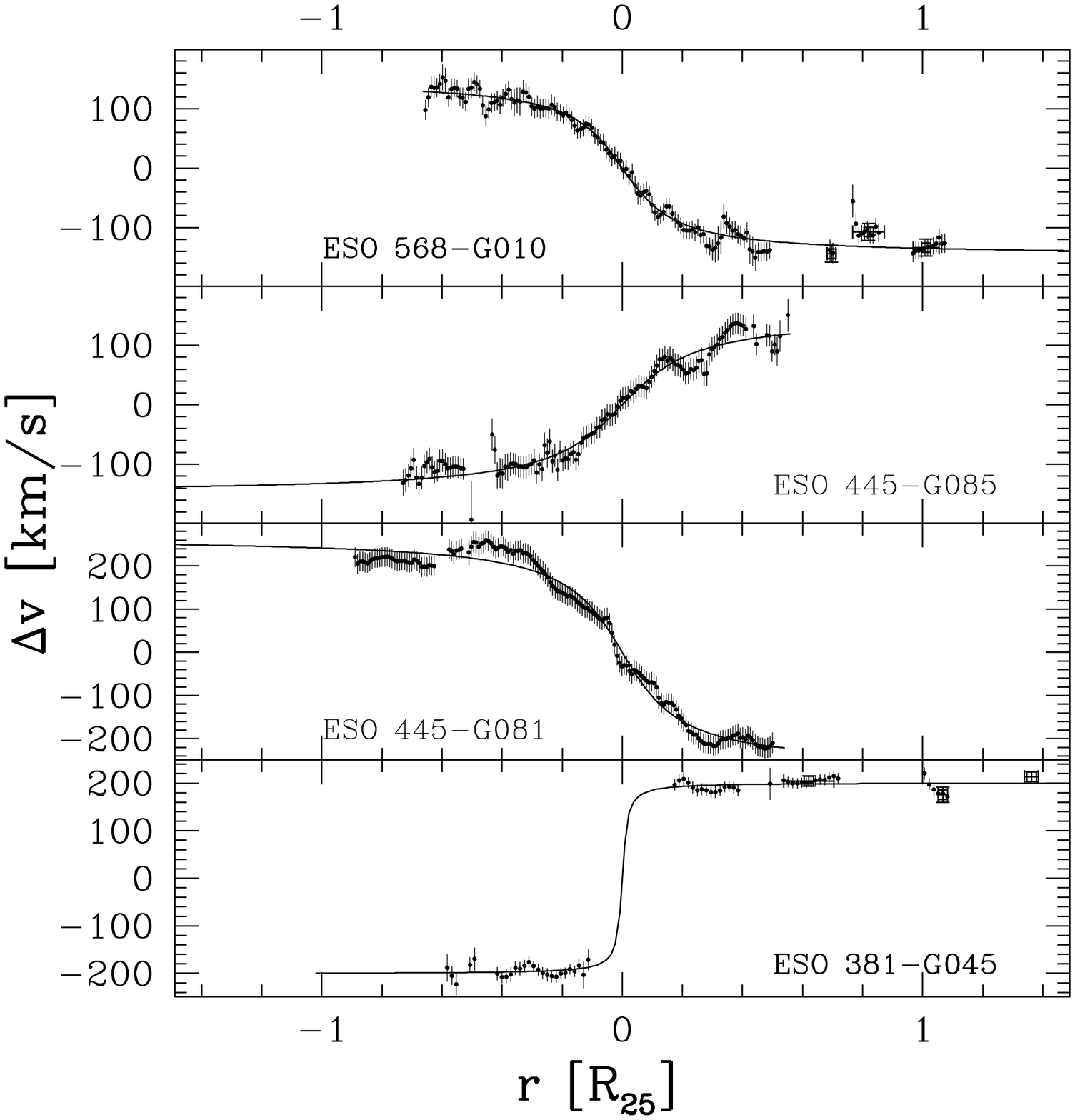}
\caption{Velocity centroids of H$\alpha$ measured relative to the systemic galaxy velocity on a row-by-row basis in the two-dimensional spectrum. The galaxies in this figure were observed with Magellan/B\&C. The solid line represents the best {\it arctan} fit. Open symbols show measurements where, to improve S/N, we have binned over larger regions of interest.
}
\label{fig_bdcomp3}
\end{figure}

\begin{figure}
%\plotone{fitrc2.eps}
\plotone{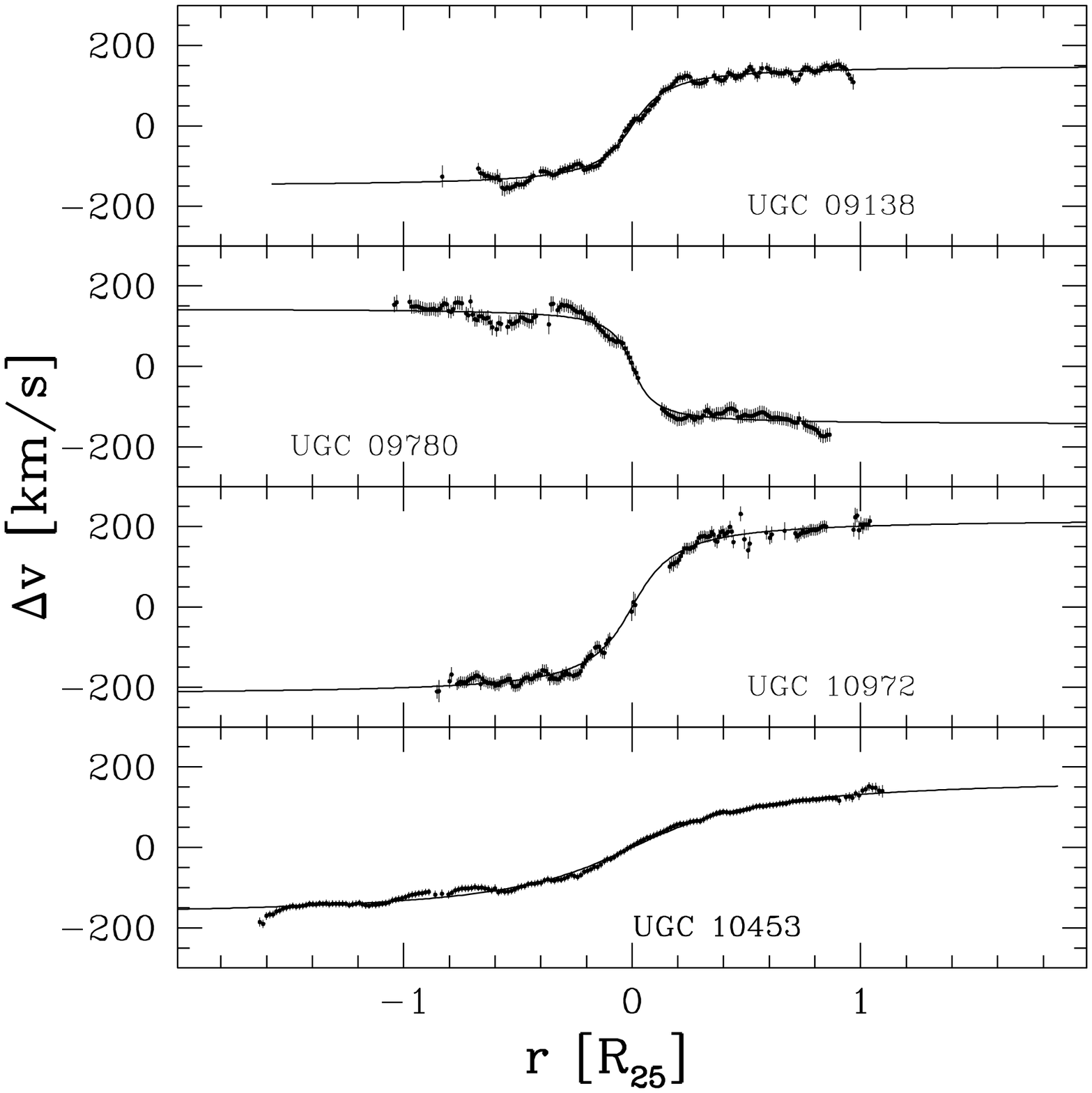}
\caption{Velocity centroids of H$\alpha$ measured relative to the systemic galaxy velocity on a row-by-row basis in the two-dimensional spectrum. The galaxies in this figure were observed with MMT/Red Channel. The solid line represents the best {\it arctan} fit. Open symbols show measurements where, to improve S/N, we have binned over larger regions of interest.
}
\label{fig_bdcomp4}
\end{figure}

\begin{figure}
%\plotone{fitrc.mcg0131.eps}
\plotone{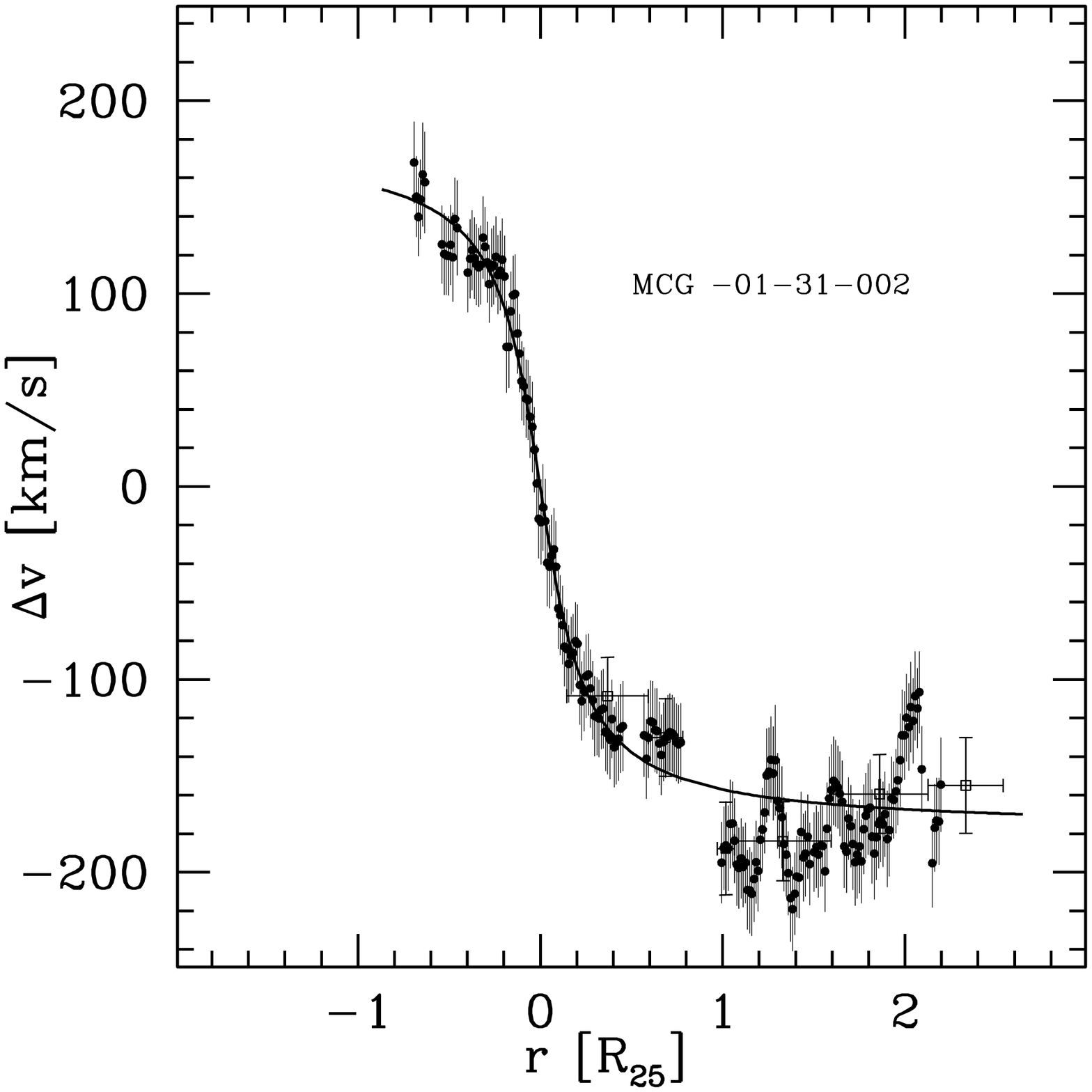}
\caption{Velocity centroids of H$\alpha$ measured relative to the systemic galaxy velocity on a row-by-row basis in the two-dimensional spectrum for MCG -01-31-002, measured at Magellan with the B\&C spectrograph. The solid line represents the best {\it arctan} fit. Open symbols show measurements where, to improve S/N, we have binned over larger regions of interest.
}
\label{fig_rcmcg0131}
\end{figure}

The rotation curves for all the galaxies in our sample are shown in Fig. \ref{fig_bdcomp}-\ref{fig_rcmcg0131} in units of the optical radius, $R_{25}$. We plot the measured line centroids for each individual row of the two-dimensional spectrum in which H$\alpha$ is detected. We investigate every outlier and comment on the interesting ones below. Fluctuations of the rotation curves around the asymptotic velocity are typically larger than the uncertainties in the individual line centroid measurements. Statistical uncertainties in the centroid are below 10 km s$^{-1}$ for regions with the strongest H$\alpha$ flux, while the dispersion about the fitted curve is $\sim 20$ km s$^{-1}$. It is possible that these small-scale fluctuations do represent actual kinematic fluctuations across the slit due to non-circular motions \citep{beauvais1999,mccluregriffith06}. However, it is also conceivable that these are due to surface brightness fluctuations across the slit. The slit image width of one arcsec is theoretically capable of introducing uncertainties in the line centroid of many tens of km s$^{-1}$, and it is therefore not possible to analyze these small-scale and small-amplitude fluctuations any further, save to say that their velocity dispersion is an upper limit on the uncertainty introduced by surface brightness fluctuations across the slit. 

We have fitted functions of the form
\begin{equation}
V(R)=V_{max}\times arctan (R/R_{0}) / (\pi/2)
\end{equation}
to the extracted rotation curves \citep{staveleysmith1990}, using a simple $\chi^{2}$ optimization with respect to the asymptotic rotation velocity $V_{max}$, the scale length $R_{0}$, the central wavelength $\lambda_{0}$, and the position of the galaxy center along the slit. Errors are a combination of the measured centroid uncertainties and the standard deviation of the data points around the fit in order to encompass the effect of surface brightness fluctuations across the slit and internal velocity fluctuations. This procedure implies that the overall quality of the fit cannot be judged from the $\chi^{2}$ value (the reduced $\chi^{2}$ will always be of order unity), but  Figures \ref{fig_bdcomp} to \ref{fig_mcg0131} confirm that this functional form describes the shape of the individual profiles well. Although this description is not a physical model, the fits provide two useful scaling parameters for comparisons, $V_{max}$ and $R_{0}$. It is remarkable that an asymptotically rising (or, in the limit of small scale lengths, flat) rotation curve describes every object in our sample well; we do not see any significant evidence for declining rotation curves at large radii. This result may arise in part because our sample is dominated by relatively faint objects \citep{catinella}, but even the brighter galaxies in our sample are consistent with an asymptotically rising or flat rotation curve out to the last measured data point. By visual inspection, only ESO 445-G 081 offers indications of a turnover. The standard deviation of data points around the fit is on the order of $\sim 10- 20$ km s$^{-1}$; this contains contributions from the centroid measurement uncertainty as well, but is dominated by slit illumation effects and small-scale velocity fluctuations in the galaxy disk.

Errors on the best-fit parameters are calculated using the method of \citet{avni76} by finding the upper and lower limits to the error contours corresponding to a $\chi^{2}$ offset of 2.331 from the best-fit $\chi^{2}$, appropriate for a fit with two degrees of freedom. We list the parameters resulting from the fit for each galaxy in Table \ref{tab_arctanfit}. The asymptotic velocities $v_{max}$ should not be taken as rotational velocities, especially in cases where our spectra have not probed the flat regime of the rotation curve (particularly ESO 380-G023). For purposes of recovering the rotational velocity, the fitting parameters should be used to calculate the velocity at a reasonably large radius, such as $R_{25}$, if a more uniform measure is desired.

\begin{deluxetable}{lrrr}
\tabletypesize{\scriptsize}
\tablecaption{Rotation curve parameters}
\tablewidth{0pt}
\tablehead{
\colhead{Galaxy}  & \colhead{$V_{max}$ [km s$^{-1}$]} & \colhead{$R_{0}$} & $\lambda_{0}$ [\AA] 
}
\startdata
ESO 201- G 0228 & $183.60 ^{+  3.27}_{-  4.90}$ & $ 0.202 ^{+0.020}_{-0.021}$ & $ 6651.46 ^{+0.05}_{-0.05}$\\
ESO 299- G 018  & $199.34 ^{+  6.70}_{- 11.71}$ & $-0.313 ^{+0.049}_{-0.043}$ & $ 6668.14 ^{+0.07}_{-0.07}$\\
IC 2058         & $126.51 ^{+ 27.60}_{- 17.39}$ & $ 0.380 ^{+0.175}_{-0.111}$ & $ 6592.98 ^{+0.09}_{-0.09}$\\
ESO 478- G 011  & $130.93 ^{+  8.23}_{-  8.23}$ & $-0.178 ^{+0.036}_{-0.051}$ & $ 6677.14 ^{+0.08}_{-0.08}$\\
ESO 323- G 033  & $205.91 ^{+  5.24}_{-  7.85}$ & $ 0.438 ^{+0.033}_{-0.041}$ & $ 6614.05 ^{+0.06}_{-0.06}$\\
ESO 380- G 023  & $369.12 ^{+  8.01}_{- 26.32}$ & $-1.091 ^{+0.111}_{-0.052}$ & $ 6623.81 ^{+0.03}_{-0.03}$\\
ESO 381- G 045  & $201.16 ^{+  3.93}_{-  4.33}$ & $ 0.012 ^{+0.010}_{-0.012}$ & $ 6718.89 ^{+0.09}_{-0.09}$\\
ESO 385- G 008  & $159.61 ^{+  3.37}_{-  5.99}$ & $-0.251 ^{+0.025}_{-0.022}$ & $ 6644.82 ^{+0.04}_{-0.04}$\\
ESO 445- G 081  & $265.84 ^{+  6.55}_{-  8.87}$ & $-0.145 ^{+0.019}_{-0.018}$ & $ 6657.58 ^{+0.15}_{-0.15}$\\
ESO 445- G 085  & $136.81 ^{+  6.24}_{-  8.13}$ & $ 0.155 ^{+0.033}_{-0.035}$ & $ 6656.30 ^{+0.09}_{-0.09}$\\
ESO 568- G 010  & $144.00 ^{+  4.78}_{-  5.16}$ & $-0.124 ^{+0.022}_{-0.021}$ & $ 6683.65 ^{+0.11}_{-0.10}$\\
IC 4393         & $202.36 ^{+  4.58}_{-  7.24}$ & $ 0.247 ^{+0.025}_{-0.031}$ & $ 6622.94 ^{+0.06}_{-0.06}$\\
MCG -01-31-002  & $177.72 ^{+ 14.09}_{- 13.09}$ & $-0.184 ^{+0.033}_{-0.050}$ & $ 6687.70 ^{+0.18}_{-0.16}$\\
UGC 09138       & $152.81 ^{+  5.11}_{-  5.08}$ & $ 0.121 ^{+0.016}_{-0.015}$ & $ 6663.59 ^{+0.07}_{-0.07}$\\
UGC 09780       & $144.60 ^{+  5.24}_{-  5.60}$ & $-0.069 ^{+0.016}_{-0.018}$ & $ 6675.87 ^{+0.08}_{-0.07}$\\
UGC 10453       & $178.18 ^{+ 10.47}_{- 10.63}$ & $ 0.430 ^{+0.066}_{-0.068}$ & $ 6657.86 ^{+0.07}_{-0.08}$\\
UGC 10972       & $219.36 ^{+  4.95}_{-  6.37}$ & $ 0.139 ^{+0.016}_{-0.015}$ & $ 6664.03 ^{+0.06}_{-0.07}$
\enddata
\tablecomments{Parameters of the individual galaxy rotation curve, as gathered from the {\it arctan} fit. Columns show the asymptotic rotational velocity, the {\it arctan} scale length $R_{0}$ in units of the isophotal radius $R_{25}$, and the central wavelength $\lambda_{0}$. The sign of $R_{0}$ only depends on the orientation of the slit relative to the sense of the rotation of the galaxy.}
\label{tab_arctanfit}
\end{deluxetable}

We do not observe significant correlations between the maximum extent of the H$\alpha$ rotation curve and other galaxy parameters, such as $B$-band luminosity or $V_{circ}$.

\subsection{Notes on Individual Objects}
\label{sec_spectra}

In this section, we briefly comment on specific observations in individual galaxies. We only list objects here that deserve special attention due to peculiarities in their H$\alpha$ rotation curve or other characteristics. A hard cut in signal/noise does not always perfectly separate  plausible from spurious detections. Genuine outer-disk H$\alpha$ emission can fall below a reasonable cut, while  spurious surface brightness fluctuations can be found with very large signal/noise (e.g., as a result of background objects, CRs, or other image defects). It is therefore mandatory to inspect every spectrum visually to verify whether a weak detection (regardless of its kinematic properties) is plausible or not. While, in some of our spectra, we set a relatively low S/N threshold of 3 to flag possible detections, we adopt relatively conservative criteria for evaluating possible detections visually. In general, we consider a detection real if its line profile width is consistent with the slit image width, regardless of the position of the velocity centroid.

\subsubsection{ESO 323-G033}

\label{sec_eso323}

This galaxy is one of the faintest ($M_{B}\approx -17.7$) and latest-type galaxies in our sample. It is also our deepest exposure. The rotation curve includes outer-disk emission from at least five separate sources. Four of these are close to the disk (within $R_{25}$, but distinct from the continuous, bright, inner disk), while the outermost one is at a radius of $\sim 1.5\times R_{25}$. These outer-disk emission regions show striking kinematic anomalies: The outermost region is not consistent with the extrapolated rotational velocity of the galaxy at this radius, but rather is consistent with the systemic velocity of the galaxy itself. The innermost emission region is only significant at a level of 2$\sigma$, and the formal uncertainties in the error centroid are large. Taken by itself, this data point would have to be disregarded, but the similarity of this detection to the four other emission regions, as well as its physical proximity to three of them, make it highly unlikely that this is spurious. The velocity centroid of this region is also displaced from the rotational velocity.

\begin{figure}
%\plotone{study.eso323.eps}
\plotone{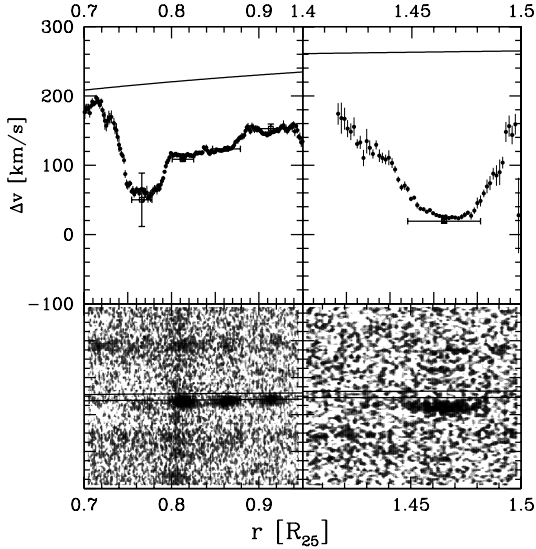}
\caption{Detailed study of Outer Disk Emission Regions in ESO 323-G033. The upper panels show the rotation curve in two different radial regions of interest (the best-fit curve is plotted as a solid line), while the bottom panels show the reduced spectra. The predicted position of slit-filling emission at the interpolated rotation velocity is indicated with two parallel, near-horizontal lines. Five distinct emission regions (the innermost one being very weak) are visible. The innermost and outermost regions deviate significantly from the extrapolated rotation velocity at their respective radii.}
\label{fig_eso323}
\end{figure}

To verify these detections and discuss whether the deviation from the mean rotation curve fit could be caused by slit illumination effects, we show in Fig. \ref{fig_eso323} a detailed study of the kinematics of these five outer-disk emission regions. The bottom panel shows a greyscale projection of the relevant parts of the spectrum, including the H$\alpha$ line as well as the neighouring [NII]6548 and [NII]6584 lines visible as horizontal structures. Solid horizontal lines indicate the limits of the slit image at the rotation velocity extrapolated from the {\it arctan} fit. The upper panels show the derived velocity relative to the systemic velocity of the galaxy. The vertical structure in the bottom left panel is continuum light from a foreground star.

All five emission regions are identifiable in this figure. The outermost emission region is evidently displaced from the extrapolated rotation velocity because the flux is almost entirely outside the region of the CCD that co-rotating gas in the slit would illuminate. For the innermost emission region, most of the flux also appears to fall well outside the relevant region.

The integrated luminosity of the outermost emission region at a distance of 36.2 Mpc is $\sim 4\times 10^{35}$ erg s$^{-1}$, about an order of magnitude fainter than M42, which has 4$\times$ 10$^{36}$ erg s$^{-1}$ \citep{wilcots}. The spatial extent along the slit is less than 2 arcsec, so that corrections for emission beyond the slit are unlikely to raise the estimate of the total luminosity by more than a factor of 2. This is considerably fainter than HII regions that are typically detectable by narrowband imaging of other galaxies, owing to the fact that optical spectroscopy can achieve a much higher S/N per pixel than imaging for detecting line emission.

The presence of gas near the systemic velocity at such large radii is surprising. One possibility is that we are observing extreme cases of projection effects, i.e., gas that is not at its maximum elongation, so that only a small fraction of the orbital velocity is projected onto the line of sight. Given the close proximity and very homogeneous appearance of these emission regions in projection, it is unlikely that they should be so widely separated along our line of sight; furthermore, attempting to explain the low line-of-sight velocity of the outermost emission region by projection of a circular rotational velocity would require the orbital radius to be so large as to be physically implausible ($\sim 11 R_{25}$). We therefore reject the hypothesis that these emission regions are moving on circular orbits in the disk plane. While it is in principle possible that at least some of these outer-disk emission regions are satellite galaxies, which would allow them to be both at large radial distances and moving on non-circular orbits, it is unlikely for even one satellite galaxy, let alone two or even five, to fall exactly on a one-arcsecond slit.

There are two remaining plausible scenarios: First, the emission
regions may be moving on strongly inclined orbits, i.e., have a
significant velocity component perpendicular to the plane of the
galaxy, which would allow them to move on nearly circular orbits.
However, this raises the question why five of these objects are seen
within 1 arcsec of the plane of the galaxy, and why three of them have
normal, disk-like kinematics. One possibility is that outer disk gas
is colliding with material on inclined orbits (a galactic fountain or
a recent accretion event may be a source of such material). In this
case, shocks may be an important source of ionization leading to the
H$\alpha$ emission. In fact, as will be discussed in a subsequent
paper, the outer-disk emission-regions exhibit anomalously high
[NII]/H$\alpha$ line ratios, which is consistent with the hypothesis
that ionization mechanisms other than local star formation are
responsible for the emission \citep{blandhawthorn}. Second, the emission regions may be on highly eccentric orbits in the disk plane. In this case, they must have acquired their unusual kinematic properties within the past several hundred million years, as elliptical orbits would lead these emitters into the dense inner disk, where they would be destroyed quickly in collisions with the rotating gaseous disk. We may therefore be seeing infalling material. 

\subsubsection{MCG -01-31-002}

\label{sec_mcg0131}

This object is an optically inconspicuous, isolated Scd galaxy of M$_{B}$=-19.6 at 5741 km s$^{-1}$, observed with the B\&C spectrograph at Magellan. Data for the inner and outer disk come from two different exposures, the one in the outer disk being much deeper than the one of the inner disk, with a small gap in coverage just short of $R_{25}$. We detect gas at extremely large radii in this object, $\sim 2.5\times R_{25}$. 

\begin{figure}
%\plotone{study.mcg0131.eps}
\plotone{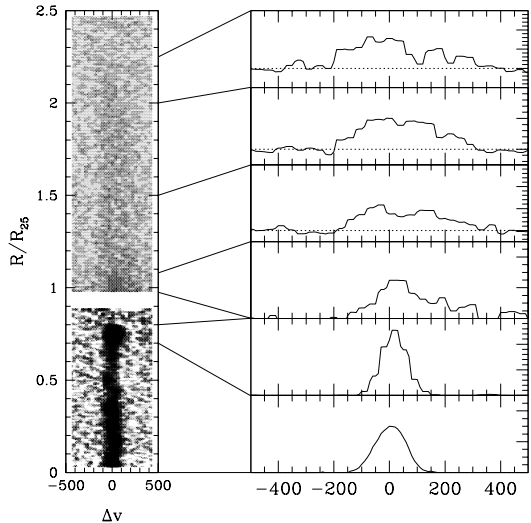}
\caption{Detailed study of the spectrum of MCG 01-31-002. The left panel shows the spectrum relative to the fitted rotation curve. Data for the inner and outer disk come from two different exposures, the one in the outer disk being much deeper than the one of the inner disk, with a small gap in coverage just short of $R_{25}$. The right panels show the line profile in various bins, whose boundaries are indicated by the lines connecting the upper to the lower panels. The most important feature is a kinematically hot component of H$\alpha$ emission, detected beyond $R_{25}$. Also notable is a bright H$\alpha$ emission region near 0.75 $R_{25}$.}
\label{fig_mcg0131}
\end{figure}

In Figure \ref{fig_mcg0131} we show the H$\alpha$ emission corrected to the mean fitted rotation curve along with cuts through the emission line at different galactic radii. The fluxes in the upper panels are scaled differently for display purposes, but the line flux in the three rightmost panels is more than an order of magnitude smaller than in the inner disk. The line profiles are fairly well-centered on the extrapolated rotation velocity and clearly following the disk rotation, with the possibility of a slightly higher-velocity tail at intermediate radii and a drop towards the outer edge. Most importantly, the line width of the faint H$\alpha$ line in the outer disk is dramatically increased, compared to the bright H$\alpha$ component in the inner disk.

Could an instrumental misconfiguration, such as a wrong focus, be responsible for this result? The slit width, estimated from the sky lines, is $2.5\pm 0.1$ pixels, while the H$\alpha$ line in the central disk is $2.67$ pixels wide. However, beyond $R_{25}$, we measure the width of H$\alpha$ to be $8.78$ pixels, corresponding to a velocity broadening of 8.4 pixels, 6.6 \AA, or $\sim 300$ km s$^{-1}$. The narrowness of the sky lines and the inner disk H$\alpha$ suggest that the instrument is working fine. Despite the apparent high velocity dispersion, the material is, on average, co-rotating with the disk. There is an exceptionally bright H$\alpha$ knot (with a surface brightness approximately three times larger than anywhere else along the slit) at $\sim 0.75\times R_{25}$. Both the anomalously high velocity dispersion and the H$\alpha$ hotspot may be signatures of on outer disk that was recently perturbed and heated by tidal interactions or accretion. It is not possible to determine whether this broadened H$\alpha$ component exists in the inner disk or is only present in the outer disk, as it is too faint to be visible on the short-term exposure centered on the galaxy center.

\subsubsection{UCG 10453}

This object, an Sd galaxy with $M_{B}=-18.25$, is remarkable for its very extended H$\alpha$ disk (out to $\sim 1.7\times R_{25}$). The outer disk emission is clumpy, suggesting outer-disk star formation as the ionization mechanism. Only one side of the galaxy was sampled with deep exposures.

\subsubsection{ESO 380- G023}

At $M_{B}\approx -17$, this Scd galaxy at $cz=2,753$ km s$^{-1}$ is among the faintest in our sample. The rotation curve is not very extended, but this object is notable for having the most-slowly rising rotation curve in our sample. Even at the largest radii to which we have sampled it ($\sim 0.9\times R_{25}$), the rotation curve shows no sign of flattening.

\subsubsection{ESO 201-G022}

This Sbc galaxy with $M_{B}\approx -19$ at $cz=4,068$ km s$^{-1}$, which is among our deepest exposures, shows H$\alpha$ emission out beyond $\sim 1.30 R_{25}$. It has a very regular rotation curve out to this radius. This galaxy is one of four targetted by \citet{degrijs} in a study of truncated disks. For this object, they find a truncation in the stellar surface brightness at $\sim 100-110$ arcsec with uncertainties on the order of 5-10 arcsec, which corresponds to $\sim 1.4 R_{25}$. Our outermost H$\alpha$ detection is therefore consistent with being at a similar radius as the stellar truncation.

\subsubsection{ESO 299-G018}

The 3.4-hour exposure with FORS1 reveals emission extending at least out to $1.33\times R_{25}$, consistent with the extrapolated rotation curve, in this Sc galaxy of M$_{B}=-18$. This object is problematic, as the H$\alpha$ emission coincides with very faint sky lines, prompting us to set a relatively high detection threshold of 7$\sigma$. 

\subsubsection{ESO 381-G045}
This Sb galaxy is the earliest-type and highest-redshift object in our sample; at M$_{B}=-19.6$, it is also fairly luminous. The continuous rotation curve of the inner disk is not particularly extended; it reaches to 0.7 R$_{25}$. However, the deep spectra reveal H$\alpha$ (at 6 and 5$\sigma$, respectively) from two point-like emission regions that are clearly detached from the inner disk, extending the rotation curve to $\sim 1.4$. The luminosities (based on the flux within the slit) are slightly higher than that calculated for the outermost emission region in ESO 323-G033; they are 10$^{36}$ and $0.6\times10^{36}$ erg s$^{-1}$. However, in this context, the reader is reminded again of the very uncertain flux calibration for the objects observed at Magellan; an uncertainty of a factor of 2 is possible in these estimates.

\subsection{Outer-Disk Detections}

\label{sec_odd}

The outermost detections of H$\alpha$ emission and their kinematics are of particular interest. If they follow the mean rotation curve, then they set a lower limit on the extent of the disk, mark the material with (probably) the highest specific angular momentum, and provide the most extreme constraints on the rotation curve at large radii. To place the tightest constraints on these quantities possible with our data, we use the following strategy: First, we determine the outermost easily detectable point of the H$\alpha$ disk, but discarding isolated detections that are not continuous for at least several arcseconds. Where in doubt about the exact edge of the emission, we typically define larger spatial bins just before and beyond the preliminary edge, and measure fluxes and velocity centroids.

We list in Table \ref{tab_outermostdet} the outermost H$\alpha$ detections on either side of the galaxy separately, and include the measured velocity relative to the interpolated rotation velocity at the bin center. In most cases, the outermost H$\alpha$ detections are around $R_{25}$, although there are individual galaxies with significantly larger H$\alpha$ disks, such as ESO 201-G022, ESO 299-G018, MCG -01-31-002, and UGC 10453, as well as galaxies with detached outer-disk emission regions (ESO 381-G045, ESO 323-G033). The velocity centroid uncertainties given in the table include contributions from surface brightness fluctuations across the slit, determined via the standard deviation around the fit as discussed before. Note that, for galaxies from the Magellan and MMT runs, the coverage is asymmetric, and deep exposures typically exist for only one side of the galaxy; the detections that are limited by radial coverage rather than by the significance threshold are marked.

\begin{figure}
%\plotone{martinstats.bin.eps}
\plotone{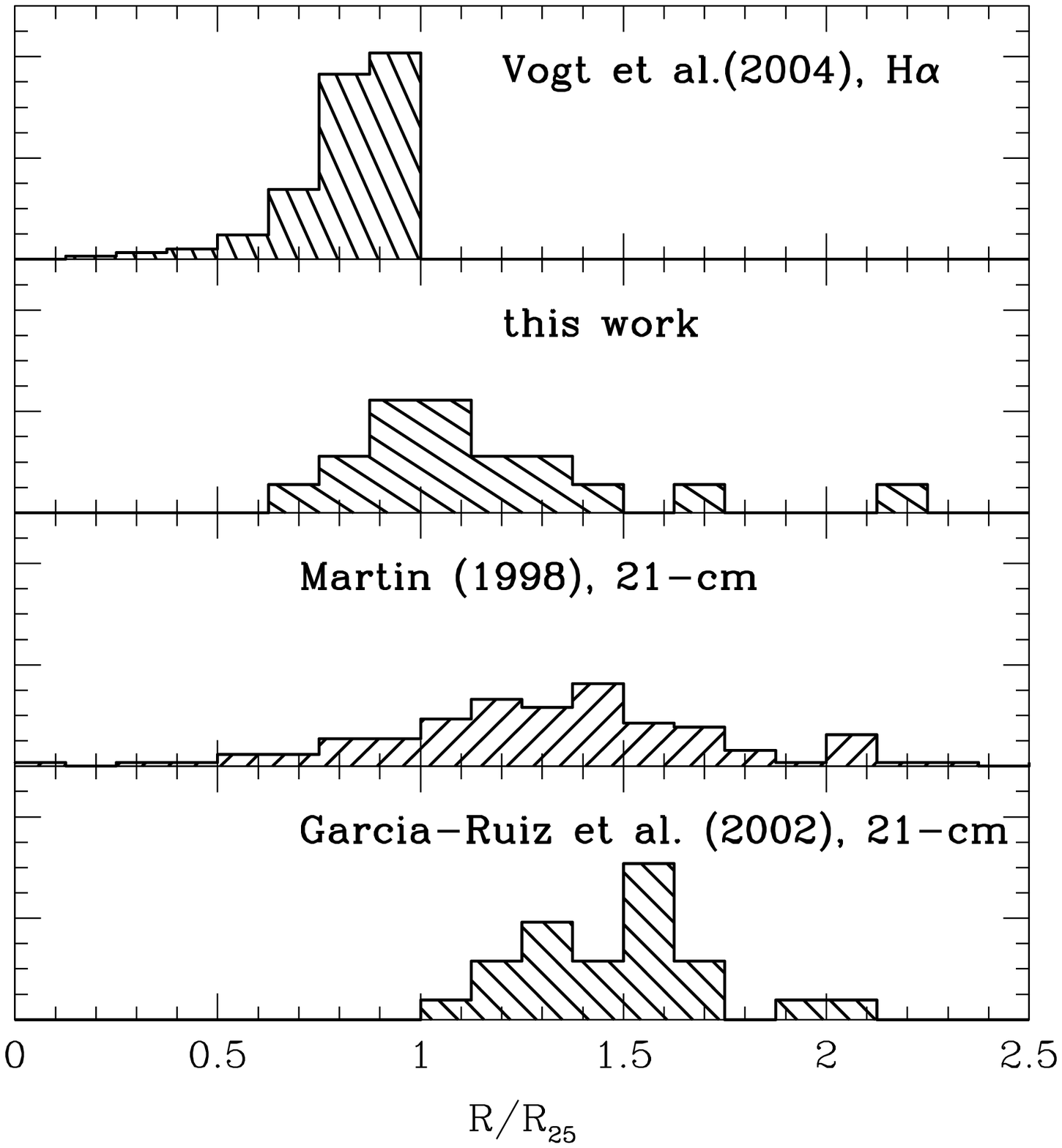}
\caption{Comparison of the maximum extent (in units of $R_{25}$) of our H$\alpha$ detections (second panel from top) to H$\alpha$ rotation curves with shorter exposure times from \citet{vogt} (top panel), a literature compilation of resolved 21-cm observations from \citet{martin1998} (third panel from top), and a 21-cm survey of late-type edge-on galaxies by \citet{garcia-ruiz} (bottom panel). Our observations probe a domain normally not observed with optical spectroscopy; there is a significant subset of galaxies with H$\alpha$ detections at radii previously only observed with radio telescopes.}
\label{fig_raddist}
\end{figure}

How well do these deep H$\alpha$ rotation curves perform in comparison to optical spectroscopy with more conventional exposure times, as well as to 21-cm studies, and are there indications that the true outer disks of galaxies are even larger than those observed here in H$\alpha$ emission? To answer these questions, Fig. \ref{fig_raddist} shows the distribution of the outermost detections in our sample in comparison to several other samples: 1) High-quality H$\alpha$ rotation curves from \citet{vogt}, using exposures of up to 3000 s duration on the Hale 200 inch telescope (itself a more powerful configuration than usual for the study of optical rotation curves). Although predominantly focused on clusters, this study also comprises a number of field galaxies, and we have selected only galaxies with morphological types later than Sb for the comparison. Instead of the $B$-band R$_{25}$ radius, this work uses the $I$-band R$_{23.5}$ radius as a reference for the optical diameter. 2) A literature compilation of resolved 21-cm observations by \citet{martin1998}. We choose a subsample of spiral galaxies of Hubble type Sb or later at cz$>$1,000 km s$^{-1}$ and require that the spatial resolution (i.e., the width of the half-power beam) be less than 20\% of the optical galaxy diameter. Although this sample is by its nature unsystematic, it provides an overview of the radii to which typical 21-cm observations probe outer disks. We use the optical radii listed in the Martin catalog itself, which are taken from the LEDA database and correspond to the the $\mu=25$ $mag$ $arcsec^{-2}$ isophotal surface brightness in the $B$-band, the same as adopted for our work. 3) A 21-cm survey by \citet{garcia-ruiz} of edge-on late-type galaxies, aimed at the study of warps, and using the Westerbork Radio Synthesis Telescope. This sample is most similar to ours with regard to selection criteria.

The optical rotation curves of \citet{vogt} generally reach to 0.75 -- 1.0 R$_{25}$. The observations recorded in \citet{martin1998} typically detect HI to radii between 1 and 1.5 times the optical diameter, although a few individual cases of extended hydrogen disks ($\sim 2$ times the optical diameter) exist. Finally, \citet{garcia-ruiz} typically detect 21-cm emission out to 1.5 R$_{25}$, although a few cases of more extended emission exist. In comparison, of the deep optical spectra presented here, many can be traced out approximately to R$_{25}$, but there is a significant number of galaxies for which H$\alpha$ detections extend considerably further, as far as 2 R$_{25}$ in one case (out of 17), a region that was previously only probed in 21-cm; more than half of our sample show detections beyond R$_{25}$. Therefore, in such cases, our observations are probing a regime that has previously not been accessed routinely with optical spectroscopy.

Figure \ref{fig_raddist} also shows that 21-cm observations on average
still probe even larger radii of the outer disk. This may have two
reasons: First, H$\alpha$ emission may simply be too faint and drop
below the detection threshold before the physical truncation of the
neutral hydrogen disk. Most ionization mechanisms (ionization by local
star formation or by UV flux from the inner disk \citep{blandhawthorn1998}) lead us to expect a declining surface brightness profile; only ionization by the cosmic ionizing background would eventually cause a flat surface brightness distribution as long as the disk is optically thick to UV radiation, but there is no indication that our observations have reached the depth required for a detection of background-ionized gas. The second possible cause is that many, and possibly most, galaxies are now thought to have warps in the outer disks \citep{sanchez-saavedra,reshetnikov,bosma,garcia-ruiz}, which means that, at large distances, the disks may physically bend away from the plane observed by the spectrographic slit. In fact, Christlein \& Bland-Hawthorn (in prep.) are currently conducting a campaign to observe warped galaxy disks with optical spectroscopy, and first obervations confirm that, in some cases, H$\alpha$ emission can be traced to larger distances in the warp than in the plane of the galaxy itself. However, we point out that the sample in the present study was specifically selected to exclude galaxies with indications of (optical) warps on publically available imaging.

We find no obvious correlation between the maximum extent of the H$\alpha$ detections and other galaxy properties such as absolute magnitude, morphology, or optical indications of warps.

What are the kinematic properties of the outermost H$\alpha$ detections? The only detection that lags in velocity significantly from the fit is the outermost one in ESO 323-G033, which has been discussed in \S \ref{sec_eso323}. There are no indications that the rotation curves turn over or decline out to the limits of the observable H$\alpha$ distribution. This implies that at least out to these radii there is no strong truncation in the halo mass profile for any of these galaxies and that the gas in the outer disk co-rotates with the inner disk and does not yet exhibit signs of transitioning to random, halo-like kinematics. ESO 323-G033 is the only exception to this second conclusion in our sample of 17.

\begin{deluxetable*}{lrrrrrr}
\tabletypesize{\scriptsize}
\tablecaption{Outermost H$\alpha$ detections.}
\tablewidth{0pt}
\tablehead{
\colhead{Galaxy}  & \colhead{R} & \colhead{$\Delta$ v } & SFB & \colhead{R} & \colhead{$\Delta$ v} & SFB \\
                  &  \colhead{[$R_{25}$]}  &  \colhead{[km s$^{-1}$]} &    &  \colhead{[$R_{25}$]} & \colhead{[km s$^{-1}$]} & 
}
\startdata
ESO 478- G 011   & $  1.20 - 1.25 $ & $   14.46 \pm   17.23 $ & $   3.88 \pm   0.69 $ & $  1.15 - 1.20 $ & $    8.99 \pm   19.12 $ & $   2.63 \pm   0.52 $\\
ESO 299- G 018   & $  1.31 - 1.34 $ & $   58.34 \pm   16.95 $ & $   3.49 \pm   0.45 $ & $  1.18 - 1.21 $ & $  -37.16 \pm   18.00 $ & $   3.79 \pm   0.51 $\\
ESO 201- G 028   & $  1.30 - 1.32 $ & $   10.00 \pm   15.19 $ & $   2.00 \pm   0.32 $ & $  1.20 - 1.26 $ & $   14.59 \pm   14.06 $ & $   2.07 \pm   0.21 $\\
IC 2058          & $  1.12 - 1.14 $ & $   20.29 \pm   30.56 $ & $   2.19 \pm   0.45 $ & $  0.94 - 0.96 $ & $  -14.53 \pm   26.74 $ & $   3.17 \pm   0.76 $\\
ESO 323- G 033   & $  1.46 - 1.49 $ & $ -136.73 \pm   12.15 $ & $   1.99 \pm   0.26 $ & $  0.75 - 0.78 $ & $  -29.02 \pm   12.14 $ & $   2.07 \pm   0.23 $\\
ESO 380- G 023   & $  0.88 - 0.90 $ & $    4.18 \pm   10.17 $ & $   2.33 \pm   0.31 $ & $  0.69 - 0.71 $ & $  -18.24 \pm   10.01 $ & $   2.11 \pm   0.34 $\\
ESO 385- G 008   & $  0.85 - 0.86 $ & $  -24.64 \pm   13.60 $ & $   8.68 \pm   1.34 $ & $  0.75 - 0.76 $ & $   21.05 \pm   12.66 $ & $  10.35 \pm   1.59 $\\
IC 4393          & $  0.95 - 0.96 $ & $  -30.24 \pm   24.93 $ & $   2.92 \pm   0.48 $ & $  0.73 - 0.74 $ & $    5.42 \pm   24.94 $ & $   3.15 \pm   0.51 $\\
ESO 445- G 081   & $  0.88 - 0.90 $ & $  -18.25 \pm   23.26 $ & $   6.03 \pm   0.98^{2} $ & $  (0.49 - 0.51)^{1} $ & $   -8.38 \pm   24.09 $ & $  80.15 \pm  11.16^{2} $\\
ESO 445- G 085   & $  0.73 - 0.75 $ & $   15.36 \pm   22.20 $ & $   6.75 \pm   1.31^{2} $ & $  (0.53 - 0.55)^{1} $ & $   35.09 \pm   27.56 $ & $  35.24 \pm   7.95^{2} $\\
ESO 381- G 045   & $  1.34 - 1.39 $ & $   13.79 \pm   10.94 $ & $   1.46 \pm   0.29^{2} $ & $  (0.57 - 0.60)^{1} $ & $   -9.79 \pm   29.09 $ & $  29.04 \pm   5.91^{2} $\\
ESO 568- G 010   & $  1.07 - 1.09 $ & $   -6.14 \pm   16.87 $ & $   3.07 \pm   0.93^{2} $ & $  (0.64 - 0.66)^{1} $ & $  -30.18 \pm   18.01 $ & $  41.40 \pm   9.26^{2} $\\
MCG -01-31-002   & $  2.14 - 2.26 $ & $  -13.66 \pm   24.40 $ & $   5.75 \pm   1.07^{2} $ & $  (0.68 - 0.70)^{1} $ & $   19.75 \pm   21.25 $ & $  57.14 \pm   8.80^{2} $\\
UGC 09138        & $  0.96 - 0.98 $ & $  -31.49 \pm   18.79 $ & $   4.14 \pm   0.91^{2} $ & $  (0.66 - 0.68)^{1} $ & $  -30.29 \pm   12.57 $ & $  26.08 \pm   5.96^{2} $\\
UGC 09780        & $  1.03 - 1.05 $ & $   15.23 \pm   19.06 $ & $   9.22 \pm   1.15^{2} $ & $  0.85 - 0.87 $ & $   32.61 \pm   17.09 $ & $   9.13 \pm   1.08^{2} $\\
UGC 10453        & $  1.57 - 1.69 $ & $   36.73 \pm   11.17 $ & $   1.24 \pm   0.35^{2} $ & $  (1.04 - 1.15)^{1}$ & $    3.84 \pm   15.26 $ & $  10.30 \pm   2.81^{2} $\\
UGC 10972        & $  1.02 - 1.04 $ & $   11.33 \pm   13.07 $ & $   3.38 \pm   0.63^{2} $ & $  0.86 - 0.87 $ & $   14.63 \pm   16.22 $ & $    2.90 \pm   0.76^{2} $
\enddata
\tablecomments{The first set of columns shows the outermost H$\alpha$ detections for each galaxy with radius, velocity relative to the interpolated rotation velocity, and H$\alpha$ surface brightness in 10$^{-18}$ erg s$^{-1}$ cm$^{-2}$ arcsec$^{-2}$. The second set shows the corresponding measurements on the opposite side of the galaxy. $^{1}$This measurement is affected by the lack of radial coverage due to the short length of the slit, rather than the surface brightness level of the H$\alpha$ emission. $^{2}$Flux calibration for runs at MMT and Magellan is uncertain.}
\label{tab_outermostdet}
\end{deluxetable*}

\subsection{Mass Density Profiles and Rotation Curves}

The existence of massive, extended dark matter halos around galaxies has been demonstrated using several lines of evidence \citep[such as satellite galaxy kinematics and weak lensing;][]{zsfw,mckay}. However, these are statistical studies, and it is therefore difficult to exclude the possibility that there exists a small minority of galaxies with distinctly different properties \citep{honma}. The kinematics of gas at large radii remains the most straightforward method of constraining the halo mass profile of individual objects because gas, if dynamically stable, must lie on closed orbits.

 We pose two questions. Is there any case among the observed rotation
 curves that is consistent with a strongly truncated dark matter halo?
 Conversely, can a truncated halo explain the few instances of
 anomalous kinematics in outer disk emission regions \citep{blandhawthorn}?

We attempt to answer these questions by first fitting a mass model that consists of an isothermal halo and a disk, to the observed rotation curves. The model has four free parameters, the halo central density and core radius, and the disk central surface density and scale length (the latter two could in principle also be constrained by the surface photometry, but we suspect that due to extinction and stellar population gradients they are unreliable as indicators of the stellar surface mass density in this sample of nearly edge-on galaxies. Once we have a best-fit model, we fix these four parameters and introduce two additional parameters, the truncation radii of the disk and halo. Normalizing the probability for the untruncated fit to unity, we reject cutoff radii for which the $\chi^{2}$ probability of the fit drops below 5\%.

We apply this analysis to the three galaxies with the most extended rotation curves: ESO 299-G018, UGC 10453, MCG -01-31-002. We also include ESO 323-G033 to verify whether a truncated halo could be responsible for the anomalous kinematic properties of at least the outermost H$\alpha$ emission region.

\begin{figure}
%\plotone{modfig.eps}
\plotone{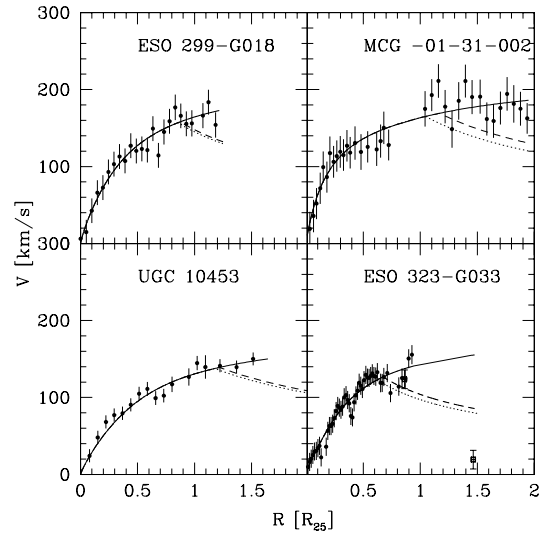}
\caption{Disk-Halo models fitted to four of our most extended rotation curves. Solid lines show the best fit for untruncated disk-isothermal halo models. The dashed line and dotted line show models for a completely truncated halo, where the dashed line is a marginally acceptable model and the dotted line a marginally rejected one. In the best cases, outer-disk rotation curves rule out halo truncations within $R_{25}$. Even a complete truncation of the halo cannot explain the anomalously low velocity of the outermost H$\alpha$ emission region in ESO 323-G033.}
\label{fig_models}
\end{figure}

We present the results in Figure \ref{fig_models}. For clarity, we show only a fraction of the measured rotation curve data points. We reach three conclusions: 1) truncations of the disk  surface mass density have subtle effects on the rotation curve and are therefore not well-constrained, 2) at least in the most extended rotation curves in our sample, dark matter halos cannot be truncated in any of these three galaxies at less than $R \sim 0.9 R_{25}$, and 3) even a dark matter halo truncated well within $R_{25}$ has difficulties reproducing the unusual kinematic properties of the outermost emission in ESO 323-G033.

\subsection{The Composite Spectrum}

\subsubsection{Construction of the Composite Spectrum}
\label{sec_constcomp}
\label{sec_composite}

All rotation curves in our sample are described well by the same analytical form; their differences can be reduced simply to differences in the characteristic scale length, $R_{0}$, and the asymptotic rotational velocity, $V_{max}$. The homogeneity of the sample provides the opportunity to combine the spectra into a ``composite spectrum''. Such a composite spectrum might allow us to detect finer and fainter details and characterize the behavior of an ``average'' galaxy.

In this section, we describe the construction of such a composite spectrum. We use the two characteristic scales obtained from the arctan fits to transform the individual galaxy spectra onto a common coordinate frame before superposing them. There are different choices available for such a common coordinate system, and each one serves a particular purpose best.

We use the radial projected distance from the galaxy center in units of $R_{25}$ as the spatial axis. For the velocity axis, we use {\bf three} mappings. First, for tracing the rotation curve to large radii and detecting emission with anomalous kinematics, we use the difference from the {\it arctan} fit, measured in km s$^{-1}$. This transformation is defined in such a manner that emission at rotational velocities greater than the interpolated rotation curve are plotted as positive and those which are less than the interpolated curve are plotted as negative. Second, we use the ratio of the observed velocity and the arctan fit. In this transformation, emission at the systemic velocity will always stack at an ordinate of $y=0$, and counterrotating gas will stack at an abscissa of $y=-1$. This mapping is therefore ideal for searching for counterrotating gas. Third, to characterize the velocity dispersion of the gas, we determine the region illuminated in the two-dimensional spectrum by gas at the exact interpolated rotation velocity that fills the slit. We then assign for any detected emission beyond that region a $y$ value that corresponds to the difference, in km s$^{-1}$, between the inferred velocity and the closest slit edge. This results in a minium possible velocity dispersion because it removes the maximal effect from uneven slit illumination.

 The flux uncertainties in the individual source pixels are calculated as outlined in Sec. \ref{sec_measure} by determining the standard deviation of flux measurements carried out in a control region that is free of line emission or absorption. For determining the uncertainties in the centroid, we apply a bootstrapping algorithm that randomly resamples the set of individual spectra from which the composite is constructed; for each resampling, the centroid is recalculated, and the standard deviation after 100 iterations adopted as the centroid uncertainty. This estimate includes uncertainties that are due to surface brightness fluctuations across the slit, and is therefore appropriate for evaluating whether there are radial trends of the velocity centroid in the composite spectrum. 

For the construction of the composite spectrum, we use all available spectra from the Gemini and VLT observing runs, and the deep, outer-disk spectra from the MMT and Magellan runs. The fact that the latter do not provide continuous spatial coverage gives rise to the concern that they could bias the shape of the surface brightness profile of the composite spectrum (a single galaxy with extremely low or high surface brightness in the MMT or Magellan samples could affect the outer parts of the composite spectrum and thus change the overall shape). However, we have tested this possibility by calculating a composite spectrum from just the seven highest-quality VLT and Gemini spectra, and see no systematic difference between both in either the overall shape of the spatial surface brightness distribution or the rotation curve, with the only exception that the H$\alpha$ emission is more extended beyond 1.3 R$_{25}$ in the composite of 17 spectra.

\subsubsection{The Extent of the Composite Spectrum}

\begin{figure}
%\plotone{profile.spatial.eps}
\plotone{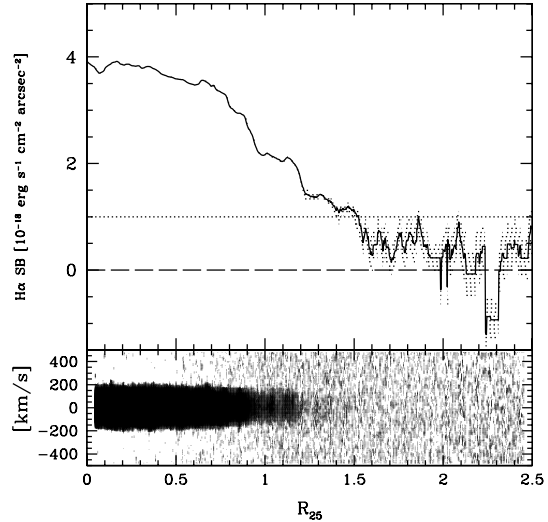}
\caption{Composite Spectrum and its surface brightness distribution. H$\alpha$ emission is clearly discernible out to $1.5\times R_{25}$ in the bottom panel. Velocities in the bottom panel are are offsets from the predicted rotation velocity. The top panel shows the H${\alpha}$ surface brightness profile. Flux scale is linear below 1, and logarithmic otherwise (see text for detailed explanation)}.
\label{fig_comspec}
\end{figure}

 The top panel of Fig. \ref{fig_comspec} shows the H$\alpha$ surface brightness as a function of radius. To include all of the data and make the low surface brightness levels legible, we plot values differently depending on the value itself as described by the following prescription:

$$y=\cases{c(\log(f/c)+1) & if $f\geq c$;\cr
f/c&if $<c$} $$

We set $c=1$, i.e., fluxes up to 1 erg s$^{-1}$ cm$^{-2}$ arcsec$^{-2}$ will be plotted linearly, while every full step above $1$ indicates an increase by a factor of 10.

H$\alpha$ emission is visible to a radius of at least $1.5 R_{25}$ in the composite spectrum. Between 1.25 R$_{25}$  and 1.5 R$_{25}$, H$\alpha$ emission is significant at over 10 $\sigma$, but it drops to 3$\sigma$ between 1.5 and 1.75 R$_{25}$, which we do not accept as a significant detection. At smaller radii, there is no doubt that the emission up to $1.5 R_{25}$ is associated with the galaxy disks. Furthermore, the spectrum very clearly shows a broad absorption feature (indicated by the lower background level) surrounding the H$\alpha$ emission within $\sim 0.75 R_{25}$. This feature is likely indicative of H$\alpha$ absorption. 

Individual galaxies exhibit a range of radii of their H$\alpha$ disks, and in particular, emission in the composite spectrum at radii around $1.5 R_{25}$ is likely to be contributed by only a small subset of galaxies in the sample. Nonetheless, even if we remove the galaxies with the most extended individual H$\alpha$ detections (ESO 201-G022, UGC 10453, MCG -01-31-002, ESO 323-G033, ESO 299-G018) emission is still clearly discernible in the composite spectrum out to $\sim1.25\times R_{25}$. In general, though, emission in the composite spectrum is not more extended than in the most extended individual contributing spectrum, suggesting that faint outer-disk emission features are not a ubiquitous phenomenon whose S/N would improve as exposures are stacked. We reserve the discussion of the H$\alpha$ surface brightness profiles, and in particular, how representative the composite is for the individual galaxies, for a subsequent paper, save for the caveat that the kinematic constraints that we gain from the outermost detections in the composite spectrum are only obtained from a subset of galaxies.

\subsection{The Composite Rotation Curve}
\label{sec_comprc}

\begin{figure}
%\plotone{tsrc.eps}
\plotone{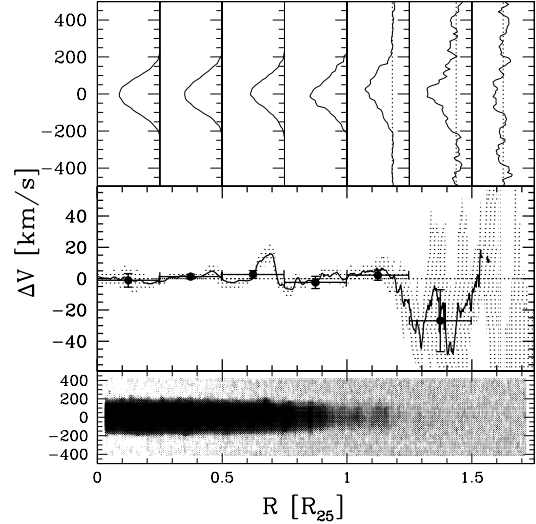}
\caption{Deviations of the Composite Spectrum from the {\it arctan} rotation curve. The bottom panel shows the rotation curve normalized composite spectrum. In the middle panel, we plot the velocity centroid of the H$\alpha$ line in the composite spectrum relative to the {\it arctan} fit (solid line, surrounded by thinner lines marking the 1$\sigma$ standard deviation of the rotation curve relative to the fit). Velocity centroids have also been measured in several wide bins and overplotted as filled circles. Finally, the top panel presents the line profiles at several radii (each profile corresponds to one of the wide-bin data points in the middle panel).
}
\label{fig_comspec_rc}
\end{figure}

We now examine the transformed two-dimensional spectrum in Figure \ref{fig_comspec} for deviations from the fitted {\it arctan} rotation curve. Flux and centroid measurements are carried out in the same way as for the individual spectra, and uncertainties are calculated as described in \S \ref{sec_composite}. The result is shown in Figure \ref{fig_comspec_rc}. Detections below 3$\sigma$ significance are not plotted. 

Measurement uncertainties of the line centroid are shown in Figure \ref{fig_comspec_rc} with a shaded area, and data points measuring the line centroid in larger bins of width 0.25 R$_{25}$ are overplotted. Out to 1.25 R$_{25}$, the data points line up extremely well with $\Delta V=0$ line, indicating that there are no systematic deviations from the {\it arctan} shape. The last significant data point between 1.25 and 1.5 R$_{25}$ exhibits a slight drop from the rotational velocity, but the uncertainties are large enough for this measurement to remain consistent with $\Delta V=0$.

The extent of this average rotation curve does not preclude the possibility that individual galaxies have more extended outer disks, however, the composite spectrum demonstrates, better than most of the individual spectra, that rotation curves of late-type galaxies follow an asymptotically rising profile, at least to $1.5 R_{25}$, and that there are no significant systematic deviations from this profile in either the positive or the negative direction. Nonetheless, it is important to be aware of a possible selection effect: the composite rotation curve at 1.5 R$_{25}$ is contributed by only a small subset of galaxies; if kinematic anomalies in the outer disk affect the maximum extent of the H$\alpha$ emission (for example, due to warps), then such anomalies may be rendered undetectable for us.

\subsection{Velocity Dispersion in the Outer Disk}

Disk galaxies consist of two kinematically very distinct regimes --- a
disk dominated by ordered, circular motions, and a halo dominated by
random motions. Are these realms strictly separated from each other,
or is there a transition region where formation mechanisms leave their
signatures in the form of an increased velocity dispersion? We have already seen possible examples of this interface region in individual spectra --- the hot outer disk in MCG -01-31-002, and the gas on anomalous orbits in ESO 323-G0033. However, the question remains whether these are general features in the galaxy population, or whether they are rare and possibly short-lived phenomena that are only observable in a small fraction of objects. In this section, we addres this question by studying the velocity dispersion of the outer-disk gas along our line of sight in the composite spectrum.

Optical spectroscopy has advantages and disadvantages for answering this particular question. Because the width of the slit is narrower than the typical beam size of 21-cm observations (even interferometric observations usually do not reach arcsecond-scale resolution), the tails of the observed line profile are representative of the velocity dispersion rather than of a gradient in the line-of-sight velocity across the field (unless the galaxy is seen in an almost perfect edge-on configuration). Its drawback is, once again, the possibility that surface brightness fluctuations across the slit might be mistaken for velocity shifts. In a composite spectrum, these shifts might be mistaken for an increased velocity dispersion.
To avoid this confusion, we use for this analysis the third type of mapping the individual spectra to the velocity axis described above, i.e., we cut out the region of each spectrum that could be covered by emission anywhere in the slit moving at the rotational velocity. Any emission beyond this region must originate from material that is not moving at the rotational velocity.

\begin{figure}
%\plotone{dispersion.eps}
\plotone{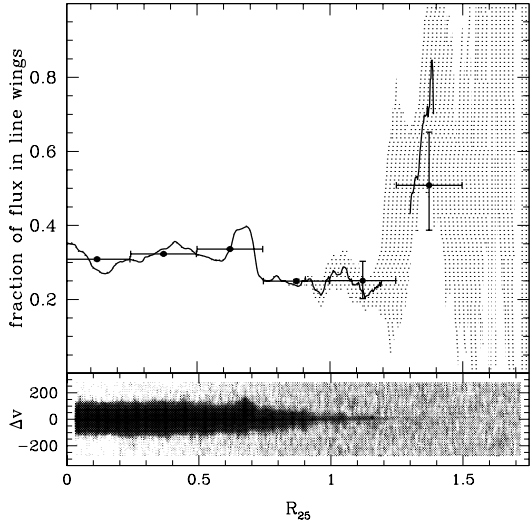}
\caption{The line-of-sight velocity dispersion of the disk gas, measured by the amount of flux beyond the predicted region of the slit image for gas co-rotating with the galaxy disk. The bottom panel shows the composite spectrum after the region of the line profile covered by the slit image has been removed from all individual spectra. Remaining flux emanates only from the wings of the observed line profile, which must be due to kinematic deviations from the fitted rotation curve. The upper panel shows the fraction of this flux compared to the total line flux. Solid lines and the shaded area show the ratio and its uncertainties measured on a row-by-row basis, while data points with error bars show the ratio measured in larger bins of width 0.25 R$_{25}$. There is no solid evidence for an increase in the velocity dispersion with radius, although the final bin beyond 1.25 $R_{25}$ is weakly constrained.
}
\label{fig_vdisp}
\end{figure}

In Figure \ref{fig_vdisp} we show the newly constructed composite spectrum after removing the emission within the projected slit image at the rotation velocity in each individual spectrum. The top panel shows the ratio of H$\alpha$ flux beyond the projected slit image relative to that within the slit image. The error bars represent two extreme cases: the upper error bar shows the ratio, assuming that the flux in the wings was underestimated by 1$\sigma$, and the total flux overestimated by 1$\sigma$. The lower error bar is calculated analogously. 

The fraction of flux in the wings beyond the slit image is consistenly in the range between 0.2 and 0.3, with no indication of an increase up to a distance of 1.25 R$_{25}$. Beyond 1.25 R$_{25}$, constraints are weaker; the last data point shows a rise, but is still consistent with the range seen in the inner disk due to the large uncertainties. Removing MCG -01-31-002, the galaxy with known high velocity dispersion, from the sample does not alter these results significantly; the recovered values shift within the measured uncertainties. We have also tested whether the flux in the line wings may be overestimated at large radii by the fact that the flux is always measured around the position of the strongest H$\alpha$ detection; this carries the risk that the flux is evaluated at the position of a spurious detection or at a position that is not even included in the original measurement of the full line flux. We have done so by a) measuring the flux at a fixed position in the composite spectrum, thereby eliminating the risk of ``hunting for significance'', and by b) increasing the width of the measurement window that we use for detecting the full line flux. The measurements show no significant deviations of the recovered wing flux fractions from the presented case. 

Finally, the flux in the wings may be overestimated at small radii if the H$\alpha$ emission follows the {\it arctan} profile better at some radii than at others. Systematic deviations from the {\it arctan} shape would manifest themselves by spilling more light beyond the calculated position of the slit image. To test this, we have constructed a new version of the composite spectrum in which the individual spectra are not aligned along the interpolated rotation curve, but rather, along the empirically-determined line centroids. Repeating the analysis on this spectrum, the wing flux fraction changes by a small amount, but typically by no more than 0.05, which does not alter the observed trend. Only the data point between 1.25 and 1.5 R$_{25}$ shifts downward by almost 0.2, but this change is not significant. Therefore, this test yields no reason to change our previous conclusion that there is no increasing trend of velocity dispersion with radius.

\subsection{A Search for Counter-Rotating Gas}

We now turn from examining the bulk behaviour of the gas to examining whether there is a minority of gas with distinctly different kinematic properties. If galaxies grow by hierarchical build-up, discrepant velocities should arise naturally as the galaxy aquires material with distinctly different angular momenta. Whereas in the dense, inner disk, such material is quickly assimilated through collisions with the gaseous disk, the lower densities and larger dynamic timescales at the outer edge of the disk will preserve such kinematic signatures. Material in counterrotation, in particular, provides striking evidence of hierarchical growth.

Again, we use the composite spectrum to maximize our sensitivity to very faint H$\alpha$ emission, and to that end, we transform the individual spectra into a coordinate frame where such emission is readily identifiable. For this purpose, we use the second mapping prescription, as described in \S \ref{sec_constcomp}, by defining the coordinate along the vertical axis as the ratio of the measured velocity to the interpolated rotational velocity. In this coordinate system, material at the systemic velocity of the galaxy will always appear at a coordinate of 0, while counter-rotating material will be found at $-1$. To quantify the H$\alpha$ emission on potentially counterrotating orbits, we integrate the flux between $-$1.5 and $-$0.5 of the rotational velocity at each radius, and divide it by the flux between 0.5 and 1.5 of the rotational velocity. 

\begin{figure}
%\plotone{counterrot.eps}
\plotone{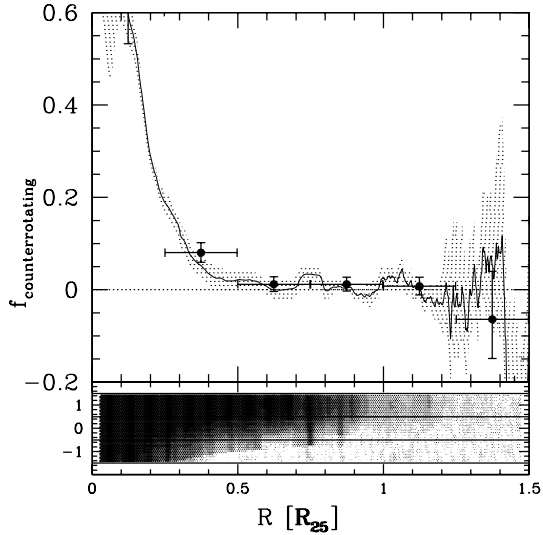}
\caption{Constraints on the mean amount of counter-rotating H$\alpha$-emitting gas in the composite spectrum. The bottom panel shows the composite spectrum, transformed  into a coordinate system where gas at the circular velocity is at the ordinate +1, while counter-rotating gas is expected at -1. The upper panel shows the fraction of gas with counter-rotating kinematics, measured in the two strips around $y=+1$ and $y=-1$ indicated in the bottom panel with horizontal lines. Within $\sim 0.4\times R_{25}$, the wings of the H$\alpha$ line profile of co-rotating gas are indistinguishable from gas on counter-rotating orbits, so that this figure should only be considered at larger radii. Beyond $0.5\times R_{25}$, the fraction of H$\alpha$ emission with negative specific angular momentum is small and consistent with zero. The solid line and shaded region show the measurements on a row-by-row basis, and data points with error bars are binned with a width of 0.25 R$_{25}$.
}
\label{fig_counterrot}
\end{figure}

In Figure \ref{fig_counterrot} we plot the resulting ratio for the composite spectrum. At small radii, within $\sim 0.4 R_{25}$, where the rotational velocities of the individual galaxies are still low, the tails of the co-rotating H$\alpha$ emission line spill into the region where we search for counter-rotating gas, and therefore, these high ratios are not indicative of counter-rotation. At larger radii (beyond $0.5 R_{25}$), the fraction of H$\alpha$ emission with negative angular momentum is consistent with zero and suggests that no more than a few percent of the H$\alpha$ emission comes from gas on such orbits. Again, similar caveats apply as previously: at radii beyond 1.25 $R_{25}$, few galaxies contribute to the composite, and if galaxies with counterrotating gas somehow have smaller H$\alpha$-bright disks, the true abundance of such kinematic features may be underestimated.

\section{Conclusions}

Based on the expectation that any accumulation of hydrogen will emit H$\alpha$ due to the omnipresence of ionizing flux (due to local star formation, star formation in nearby galaxies, AGN, or the metagalactic UV flux), we obtained deep optical long-slit spectroscopic observations to detect H$\alpha$ emission and to study the properties of the outer disks of galaxies. The purpose of this campaign was to characterize the kinematic behaviour of the outer disk gas and star forming regions, and in particular to search for kinematic anomalies in galaxies beyond the stellar disk that could be signatures of disturbances due to tidal interactions or mass accretion and might be the radial demarcation between galaxy disk and halo.
We detect H$\alpha$ emission down to fluxes of $\sim$ several $10^{-19}$ erg erg s$^{-1}$ cm$^{-2}$ arcsec$^{-2}$, and present in this paper our findings regarding the extent and kinematics of H$\alpha$-emitting gas in the outer disk. Further papers will use the same spectra to study the H$\alpha$ and stellar surface brightness profiles as well as the metallicity of outer-disk gas.

From our observations and analysis, we conclude for late-type, sub-L$^{*}$ galaxies as a class that:

\begin{itemize}
\item Deep optical spectroscopy is a powerful complement to 21-cm radio observations for studies of the outer gaseous disks of galaxies. Deep optical spectroscopy offers an good combination of sensitivity, spatial resolution, and kinematic resolution, and, in the best cases, probes the baryonic disk to similar radii as typical 21-cm maps. In our composite spectrum, we trace H$\alpha$ emission to 1.5 $R_{25}$ (measured edge-on), and in individual galaxies, we reach $\sim 2 R_{25}$ in some cases, comparable to the largest radii traced in 21-cm studies. 

\item There is a great variety in the distribution and extent of H$\alpha$ emission among outer disks. In at least one case, spatially continuous H$\alpha$ emission extends to beyond 2$R_{25}$, while in others, outer-disk H$\alpha$ emission is restricted to individual regions. Even in some of our deepest spectra, H$\alpha$ is completely absent beyond $R_{25}$. With our sample of 17 late-type objects, we are unable to relate this pattern to other galaxy properties, and it is unclear whether the lack of H$\alpha$ emission at large radii in some galaxies is due to a lack of ionizing flux or to an actual truncation or warp in the gaseous disk itself.

\item The kinematics of the outer disks are generally disk-like, with flat rotation curves and small velocity dispersion, out to the outermost edge of the detected H$\alpha$, typically up to $1.5 R_{25}$. The rotation curves in this sample rise asymptotically without major breaks or systematic deviations.

\item There is no evidence for an increase in the velocity dispersion towards larger radii, as measured by the fraction of the H$\alpha$ flux in the high-velocity tails of the H$\alpha$ line. Since our survey is only sensitive to gas in the plane of the galaxy, deviations from the extrapolated rotation curve that may be associated with outer-disk warps may not be detected in this search; a follow-up program to study the kinematics of gas in warps is currently ongoing.

\item There is little, if any, counter-rotating gas at large radii. We place an upper limit of a few percent on such gas (under the prior that only gas that is in the plane of the galaxy and irradiated by a sufficiently strong ionizing flux is detectable to us). This result could mean that in these predominantly late-type galaxies tidal interactions or mergers that could give rise to counter-rotation are rare or that, even at these large radii, gas densities are sufficiently large to quickly eradicate gas moving on such anomalous orbits. 

\end{itemize}

In constrast to the general trends, we find several objects that are notable for anomalous kinematics in the outer disk:

\begin{itemize}
\item ESO 323-G033, a very late-type, faint object, exhibits several small, but homogeneous outer-disk H$\alpha$ emission regions with luminosities as low as $\sim 10^{36}$ erg s$^{-1}$. Two of these have velocities close to the systemic velocity of the galaxy, rather than the expected circular velocity at the respective radii. The similar appearance of all these emission regions, as well as their close proximity in projection, make it unlikely that projection effects alone are responsible for the low line-of-sight velocities. 
It is more likely that they are either moving on eccentric orbits in the plane, which implies that they cannot be old, or on orbits perpendicular to the plane, in which case it is likely that they are being highlighted as they impact gas in the outer disk.
\item MCG -01-31-002 exhibits an outer disk with an exceptionally large velocity dispersion, on the order of the circular velocity, but which is nevertheless co-rotating. 
\end{itemize}

These unusual kinematic featurs may join other phenomena, such as stellar streams observed in our own Milky Way \citep{helmi} and around other, nearby galaxies \citep{ferguson2002}, as observational indicators of hierarchical growth.

While many of the galaxies in our sample show regular, disk-like kinematics as far as we can trace the H$\alpha$ emission, the fact that at least 2 out of 17 objects show anomalous kinematics in the outer disk is particularly noteworthy, given that our targets were selected to be isolated late-type galaxies with no obvious signs of perturbations or interactions. The true fraction of objects with such anomalies may lie even higher, as even our deepest spectra do not necessarily provide a complete census of the hydrogen in the outer disk. 

\acknowledgements

Based on observations obtained at the Gemini Observatory (program ID GS-2005A-C-4), which is operated by the
Association of Universities for Research in Astronomy, Inc., under a cooperative agreement
with the NSF on behalf of the Gemini partnership: the National Science Foundation (United
States), the Particle Physics and Astronomy Research Council (United Kingdom), the
National Research Council (Canada), CONICYT (Chile), the Australian Research Council
(Australia), CNPq (Brazil) and CONICET (Argentina).

 Based on observations made with ESO Telescopes at the Paranal
 Observatories under programme ID $<$074.B-0461$>$. This paper includes data gathered with the 6.5 meter Magellan Telescopes
located at Las Campanas Observatory, Chile.
Observations reported here were obtained at the MMT Observatory, a joint
facility of the University of Arizona and the Smithsonian Institution.

This research has made use of the NASA/IPAC Extragalactic Database (NED) which is operated by the Jet Propulsion Laboratory, California Institute of Technology, under contract with the National Aeronautics and Space Administration.

We thank Joss Bland-Hawthorn, Michael Pohlen, and Marc Verheijen for discussions that helped shape the content of this paper, and the anonymous referee for very helpful comments that molded it into its final form.

D.C. gratefully acknowledges financial support from the Fundaci\'on Andes. DZ acknowledges that this research was supported in part by the National Science Foundation under Grant No. PHY99-07949 during his visit to KITP, a Guggenheim fellowship, generous support from the NYU Physics department and Center for Cosmology and Particle Physics during his sabbatical there, NASA LTSA grant 04-0000-0041, NSF AST-0307482, and the David and Lucile Packard Foundation.

%{\it Facilities:} \facility{Magellan:Clay ()}, \facility{Gemini:South ()}, \facility{VLT:Kueyen ()}, \facility{MMT ()}

\end{document}